\documentclass[journal]{IEEEtran}
\usepackage[T1]{fontenc}

\usepackage{xcolor}
\usepackage{cite}
\usepackage{amsfonts}
\usepackage{amssymb}
\usepackage{amsmath}
\usepackage{amsthm}
\usepackage{bm}
\usepackage{algorithm}
\usepackage{algorithmic}
\usepackage{array}
\usepackage{booktabs} 
\usepackage{amsmath}

\usepackage{multirow}
\usepackage{balance}  
\usepackage{subfigure}
\usepackage{textcomp}
\usepackage{stfloats}  
\usepackage{url}
\usepackage{verbatim}  
\usepackage{graphicx}
\usepackage{epstopdf}
\usepackage{graphicx}

\usepackage{subfigure}

\usepackage[colorlinks=false,
linkcolor=blue,
anchorcolor=blue,
citecolor=blue]{hyperref}

\begin{document}

\title{AMBER: An Adaptive Multimodal Mask Transformer for Beam Prediction with Missing Modalities}

\author{Chenyiming~Wen,~\IEEEmembership{Graduate Student Member,~IEEE,}
        Binpu Shi,~\IEEEmembership{Student Member,~IEEE,}
        Min~Li,~\IEEEmembership{Member,~IEEE,} 
        Ming-Min Zhao,~\IEEEmembership{Senior Member,~IEEE,}
        Min-Jian Zhao,~\IEEEmembership{Member,~IEEE}
        and~Jiangzhou Wang,~\IEEEmembership{Fellow,~IEEE}
\thanks{Chenyiming Wen, Binpu Shi, Min Li, Ming-Min Zhao, and Min-Jian Zhao are with  the College of Information Science and Electronic Engineering and the Zhejiang Provincial Key Laboratory of Multi-Modal Communication Networks and Intelligent Information Processing, Zhejiang University, Hangzhou 310027, China (e-mail: cymwen@zju.edu.cn; bp.shi@zju.edu.cn;  min.li@zju.edu.cn; zmmblack@zju.edu.cn; mjzhao@zju.edu.cn). 

Jiangzhou Wang is with the National Mobile Communications Research Laboratory, Southeast University, Nanjing 210096, China, and also with Purple Mountain Laboratories, Nanjing 211111,
China.(e-mail: j.z.wang@kent.ac.uk).

}
}

\markboth{}%
{Shell \MakeLowercase{\textit{et al.}}: Bare Demo of IEEEtran.cls for IEEE Journals}

\maketitle
\begin{abstract}
With the widespread adoption of millimeter-wave (mmWave) massive multi-input-multi-output (MIMO) in vehicular networks, accurate beam prediction and alignment have become critical  for high-speed data transmission and reliable access. While  traditional beam prediction approaches primarily rely on in-band beam training, recent advances  have started to explore  multimodal sensing to extract environmental semantics for enhanced prediction. However, the performance of existing multimodal fusion methods degrades significantly in real-world settings because they are vulnerable to missing data caused by  sensor blockage, poor lighting, or GPS dropouts.
To address this challenge, we propose AMBER ({A}daptive multimodal {M}ask transformer for {BE}am p{R}ediction), a novel end-to-end framework that processes temporal sequences of 
image, LiDAR, radar, and GPS data, while adaptively handling arbitrary missing-modality  cases. AMBER introduces learnable modality tokens and a missing-modality-aware mask to prevent cross-modal noise propagation, along with a learnable fusion token and multihead attention to achieve robust  modality-specific information distillation and feature-level fusion. Furthermore, a class-former-aided modality alignment (CMA) module and temporal-aware positional embedding are incorporated to preserve temporal coherence and ensure semantic alignment across modalities, facilitating   the learning of modality-invariant and temporally consistent representations for beam prediction. Extensive experiments on the real-world DeepSense6G dataset demonstrate that AMBER significantly outperforms existing  multimodal learning baselines. In particular, it maintains high beam prediction accuracy and robustness even under severe missing-modality scenarios, validating its effectiveness and practical applicability.

\end{abstract}


\IEEEpeerreviewmaketitle

\section{Introduction}
Vehicular  communication network (VCN) has emerged as a promising technology for intelligent transportation systems, enabling applications such as autonomous driving, real-time traffic management, and cooperative safety services \cite{cheng2022integrated,liu2022survey,tan2024beam}. The stringent  data rate and latency requirements of these applications make millimeter-wave (mmWave) communication a crucial technology for VCNs. Indeed, mmWave has been recognized as a cornerstone for six-generation (6G) wireless networks, owing to its abundant spectrum resources and its capability to meet the high-throughput, low-latency demands of vehicle-to-infrastructure (V2I) communications \cite{mishra2019toward}.
However, mmWave signals suffer from severe  path loss and strong sensitivity to blockages from vehicles, buildings, and roadside infrastructure, posing significant challenges to link reliability and coverage.  To mitigate these issues,  beamforming and beam training, enabled by large-scale multiple-input multiple-output (MIMO) antenna arrays at base stations (BSs), have become essential for mmWave-based V2I systems, where rapid beam alignment is critical due to sub-second variations in the optimal beam direction under high mobility.

Traditional beam training approaches, such as exhaustive search and hierarchical search, have been extensively studied  to establish reliable V2I links \cite{li2019explore,liu2017millimeter,9925255,gao2020estimating,liu2025adaptive}. Exhaustive search  determines the best beam by sweeping all candidate beams in the codebook and selecting the one that achieves  the highest  signal-to-noise ratio (SNR). While accurate, this method incurs excessive training overhead. Hierarchical search reduces the overhead by progressively narrowing the beam search across multiple codebook levels \cite{xiao2016hierarchical}. However, both approaches   
suffer from high beam training overhead, especially in high-mobility scenarios.

Meanwhile,  deep learning (DL) has been explored for data-driven beam prediction, aiming to reduce or eliminate explicit beam training \cite{morais2023position,xu2023multi,al2022intelligent,jiang2022lidar,demirhan2022radar}. Early  works predominantly adopted single-modal learning, leveraging information from one sensor type, a paradigm that fails to exploit the complementary strengths of diverse sensing modalities. 
For example, GPS coordinates were used in \cite{morais2023position} to infer beam directions, leading to  overhead savings.
Reference  \cite{xu2023multi} proposed a vision-aided scheme that combines  3D object detection and user matching to estimate location distributions, while \cite{al2022intelligent} utilized computer vision for blockage prediction and proactive handover, achieving a 40\% quality-of-service (QoS) improvement. 
Other studies leveraged LiDAR \cite{jiang2022lidar} or radar \cite{demirhan2022radar}  data using a gated recurrent unit for LiDAR features and convolutional neural networks (CNNs) for radar feature maps (including range-angle and range-velocity maps). Despite their benefits, single-modality methods lack robustness: visual data degrades under occlusion or low-light conditions, whereas radar provides limited semantic context.

To improve robustness, multi-modal fusion has emerged as a powerful technique for robust beam prediction by combining complementary information from multiple sensors such as cameras, LiDAR, radar, and GPS \cite{11133431,11418645,charan2022vision,zhu2025advancing,tian2023multimodal,shi2024multimodal,zhang2024integrated,nazar2025enwar}. 
For instance,  \cite{charan2022vision} extracted visual features via a residual network (ResNet) and fused them with positional information through a multilayer perceptron (MLP) for beam prediction, demonstrating improved performance compared to single-modality methods. The cross-modal feature enhancement framework in  \cite{zhu2025advancing} instead jointly processed image and radar inputs using an uncertainty-aware dynamic fusion module and leveraged temporal correlations across multiple frames, which resulted in superior performance to earlier fusion approaches. 
Deeper multimodal integration has since been pursued: \cite{shi2024multimodal}
combined feature vectors extracted from radar, camera, LiDAR, and GPS data, with historical beam indices, and learned their temporal dependencies using an long short-term memory (LSTM) network; \cite{10577431} proposed  a hybrid deep quantum-transformer network (QTN) for beam prediction; \cite{tian2023multimodal} proposed a multimodal transformer architecture to fuse intermediate-scale features from multiple sensors to better capture intermodal relations and temporal patterns;  \cite{shi2025bemamba} introduced a lightweight multimodal fusion framework that exploits the linear complexity of state space models; and \cite{nazar2025enwar}  introduced ENWAR 2.0  through task-oriented explainability metrics, showing that its adaptive RAG-enabled LLM framework improves beam interpretation and environmental perception in terms of faithfulness, correctness, and answer relevancy.
Although effective, all these works  assumed full modality availability, an assumption often violated in practice due to sensor blockage, adverse lighting, weather conditions, or GPS outages. Consequently, their performance degrades severely when one or more modalities are missing.

Despite its practical importance, the missing-modality problem remains relatively underexplored.
Very recent work \cite{yao2025robust} attempted to address this issue using data augmentation via random masking, modality imputation, and channel attention. However, its imputation heavily depends on training data statistics, and the network does not inherently learn to operate under arbitrary missing-modality patterns. Consequently, its generalization and robustness remain limited. These limitations motivate the development of a more general framework capable of adaptively handling missing modalities without requiring architectural modifications to the prediction network. 

In light of the above discussions, in this paper, we aim to address the following three key challenges: {(\textit{i}) How to jointly learn modality-shared and modality-specific information while maintaining robustness to arbitrary missing modalities? (\textit{ii}) How to effectively exploit temporal correlations to improve prediction accuracy and robustness? 
(\textit{iii}) How to effectively align multimodal representations to prevent feature collapse? }

To answer these questions, we propose AMBER ($\bf{A}$daptive multimodal $\bf{M}$ask transformer for $\bf{BE}$am p$\bf{R}$ediction), a novel end-to-end framework that handles arbitrary missing modalities without modifying the model structure, while maintaining temporal coherence and semantic alignment. 
The main contributions are summarized as follows: 
\begin{itemize}
  \item  First, we propose AMBER, which replaces missing inputs with learnable modality tokens to maintain a consistent input structure, and incorporates a learnable fusion token to facilitate robust feature-level fusion. In addition, a missing-modality-aware attention mask is embedded into the multihead attention module to suppress cross-modal noise and enable modality-specific knowledge distillation. This design allows AMBER to jointly capture both modality-shared and modality-specific representations, effectively overcoming the limitations of existing methods that assume full modality availability.
  \item Second, to further enhance AMBER's capability, we propose  a class-former-aided modality alignment (CMA) module for semantic alignment. Specifically, this module  projects modality features into a shared class space and extracts class-level semantic representations. By pulling matched class features closer and pushing mismatched ones apart, CMA promotes modality-invariant representations and prevents feature collapse.
  
  \item Third, to capture short-term dynamics and long-range dependencies, temporal-aware positional embeddings are added before feeding multimodal sequences into the transformer, improving prediction accuracy and stability.
  \item Finally, extensive simulation results on the real-world DeepSense6G dataset demonstrate that AMBER achieves state-of-the-art beam prediction performance, and exhibits strong robustness under diverse missing-modality conditions.
  \end{itemize}

\begin{figure}
  \includegraphics[width=0.5\textwidth]{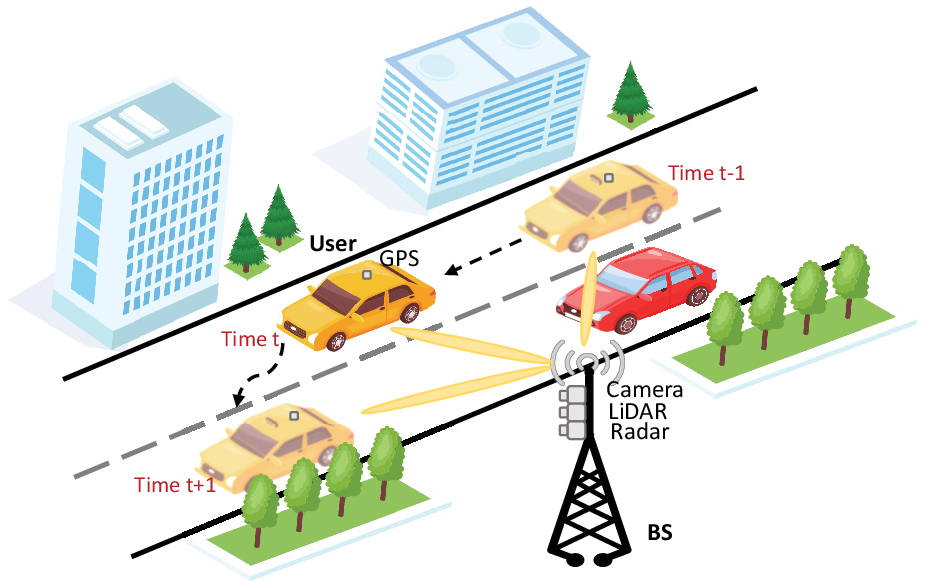}
\caption{An illustration of the system model.}\label{fig_system_model}
\end{figure}

The remainder of the paper is organized as follows. Section \ref{sec_system_model} presents the system model and beam prediction formulation. Section \ref{sec_miss_modality_trans} details the proposed AMBER framework. Section \ref{sec_simulation} provides the experimental setup and performance evaluation. Section \ref{sec_conclusion} concludes the paper.

\textit{Notations}: Scalars, vectors, and matrices are denoted by ower/upper case, bold-face lower-case, and bold-face upper-case letters, respectively. The superscript $(\cdot)^H$ and  $(\cdot)^T$ represent the Hermitian and transpose operations, respectively. The symbol  $|\cdot |$ represents the magnitude of a scalar. The symbol $\odot$ denotes element-wise multiplication. The indicator function $\bf{1}\{ \cdot\} $ returns 1 when its condition  is satisfied, and 0 otherwise. The sets $\mathbb{R}$ and $\mathbb{C}$ denote the real and complex numbers, respectively. 
Operators $\text{min}(\cdot)$ and $\text{max}(\cdot)$ return the smallest and largest values of their arguments, respectively.
$\text{FFT}_{\text{2D}}$ denotes the 2D-FFT operation. The squared $L_2$ norm of a vector ${\bf{x}}$ is written as $\| {\bf{x}} \|_2^2 = \sum_i x_i^2 $. The operator $\text{Concat}(\cdot)$  defines tokens concatenating along the feature dimension, forming a single composite input token.
The sigmoid function is defined as $\text{Sig}(x) = \frac{1}{1+e^{-x}} $.
The softmax function maps a vector  ${\bf{x}}$  to a probability distribution according to $\text{softmax}({\bf{x}})_i = \frac{\text{exp}({{x}_i})}{\sum_{i} \text{exp}({{x}}_i)} $, where $(\cdot)_i$ denotes the $i$-th element. The operator $\text{exp}(\cdot)$ denotes the exponential function with base $e$, i.e., $\text{exp}(x) = e^x$.

\section{System Model and Problem Formulation  } \label{sec_system_model}
\subsection{System Model}
Consider a mmWave V2I  system, as shown in Fig. \ref{fig_system_model}, where a BS, equipped with $N$ antennas, wishes to establish downlink connection with a single-antenna user equipment (UE). The BS employs beamforming technique and integrates multiple  sensing modalities, including camera, radar and LiDAR for environmental perception and historical scene monitoring. The mobile UE is equipped with a GPS receiver that provides real-time position estimates, which are relayed to the BS via a sub-6 GHz control link \cite{alkhateeb2023deepsense}. 
At time instance $t$, the BS might acquire the following modality data:
\begin{itemize}
  \item ${\bf{X}}_{\text{I}}[t] \in \mathbb{R}^{W_I \times H_I \times C_I}$: image data, where $W_I, H_I$ and  $C_I$ denote  the width, height and  number of channels, respectively.
  \item ${\bf{X}}_{\text{L}}[t] \in \mathbb{R}^{N_L \times 3} $: LiDAR point cloud data, where each row represents the 3D coordinate  $(x,y,z)$ of a detected point. 
  \item ${\bf{X}}_{\text{R}}[t] \in \mathbb{R}^{M_R \times S_R \times A_R} $: radar data, where $M_R$, $S_R$ and $A_R$ represent the number of  receiving  antennas, sampling per chirp, and chirp count, respectively.
  \item ${\bf{X}}_{\text{G}}[t] \in \mathbb{R}^{2 \times 1} $: GPS data, providing  the latitude and longitude coordinates of the UE.
\end{itemize}

Let ${\bf{f}}_{k} \in \mathbb{C}^{N \times 1}$ denote the $k$-th beamforming vector from a predefined codebook $\mathcal{F} = \{ {\bf{f}}_k \}_{k=1}^K$, where $K$ is the number of candidate beams. During downlink transmission, the received singal at the UE at time instance $t$ is given by 
\begin{align}
  y[t] = {\bf{h}}[t]^H {\bf{f}}_{k[t]} s + n[t],
\end{align}
where ${\bf{h}}[t] \in \mathbb{C}^{N \times 1}$ is the mmWave channel vector between the BS and UE, $s$ is the transmitted signal that satisfies power constraint $|s|^2 = 1$, and $n[t] \sim \mathcal{CN}(0,\sigma^2)$ is the complex Gaussian white noise with variance  $\sigma^2$. 

It is well established  that mmWave channels are sparse, typically comprising only a few dominant propagation paths. 
As a result, highly directional beamforming is required to achieve sufficient link gain. To ensure practical relevance and robustness, all evaluations in this work are conducted using a real-world, publicly available dataset \cite{alkhateeb2023deepsense}, rather than synthetic channel models.

\subsection{Problem Formulation}

The objective of beam prediction is to identify the optimal transmit beam that maximizes the effective channel gain, i.e., 
\begin{align}
  \label{eq:optimal_beam}
  {{k}^*[t]} =\underset{k[t] \in \{1,2, \ldots, K\}}{\arg\max} \, \left| {\bf{h}}[t]^H {\bf{f}}_{{k}[t]} \right|^2.
\end{align}

\par Traditional pilot-based beam training can achieve this objective, but it becomes prohibitively costly for large codebooks with narrow beams, as a substantial number of pilots are required. Since the optimal beam direction strongly depends on the geometric relationship among the BS, UE, and surrounding environment, DL-based beam prediction offers a promising alternative by levering  multimodal sensing information. The BS collects heterogeneous sensor data, including camera images, LiDAR point clouds, radar reflections, and positional coordinates as mentioned above, to provide rich contextual cues for learning environment-beam relationships. However, practical factors such as sensor blockage, adverse lighting, or weather conditions may render some  modalities unavailable. To model this, let $\mathcal{I}=\{I,L,R,B,G\}$ denote the set of sensing
and communication modalities, corresponding to image, LiDAR, radar,
beam, and GPS, respectively, with cardinality number I. For each modality $i\in\mathcal{I}$, we define an availability
indicator $m_i\in\{0,1\}$, where $m_i=1$ indicates that the
corresponding modality is available and $m_i=0$ otherwise.
The modality-availability vector is then expressed as
\begin{align}\label{eq_modality_avail}
{\bf m}=[m_I,m_L,m_R,m_B,m_G]\in\{0,1\}^5.
\end{align}

To capture both spatial and temporal correlations, the BS collects historical multimodal observations over a sliding window of length $W$.
For notational simplicity, we define
$\mathbf{X}_{j}^{W}[t]\triangleq \{X_j[\tau]\}_{\tau=t-W+1}^{t}$ for
$j\in\{I,L,R\}$ and $\mathbf{X}_{G}^{W}[t]\triangleq \{X_G[\tau]\}_{\tau=t-1}^{t}$. In addition,  historical  beam indices
$\mathbf{X}_{B}^{W}[t-1]\triangleq \{X_B[\tau]\}_{\tau=t-W+1}^{t-1}$ also serve  as side information and are readily available at the BS to facilitate beam prediction.

Our goal  is to learn an effective mapping function ${\mathcal{P}}(\cdot ; \bf{\Theta})$, parameterized by $\bf{\Theta}$, that predicts the optimal beam index ${\hat{k}^*}[t]$ under  arbitrary missing-modality conditions, while minimizing training overhead. Formally, we aim to design 
\begin{equation}
  \hat{k}^{*}[t]
  \!=\!
  \mathcal{P}\!\left(
  \mathbf{X}_{I}^{W}[t],
  \mathbf{X}_{L}^{W}[t],
  \mathbf{X}_{R}^{W}[t],
  \mathbf{X}_{G}^{W}[t],
  \mathbf{X}_{B}^{W}[t-1],
  \mathbf{m}[t];
  \bf{\Theta}
  \!\right).
  \label{eq:beam_prediction}
  \end{equation}

\section{Robust Multimodal Beam Prediction With Missing Modalities} 
\label{sec_miss_modality_trans}
Note that each sensing modality provides  complementary information for beam prediction.
The image modality offers  rich semantic cues but is vulnerable to occlusions and adverse weather. LiDAR offers high-resolution 3D spatial information, though its point density decreases at long ranges. Radar provides robust range and velocity measurements under challenging lighting or weather conditions, but with limited angular resolution. GPS delivers reliable positional information, yet lacks fine-grained environmental detail. Fusing these heterogeneous modalities allows the system to exploit their respective strengths, enabling one modality to compensate for the limitations or failures of another and thereby improving overall prediction robustness.

\begin{figure*}[t]
  \includegraphics[width=0.98\textwidth]{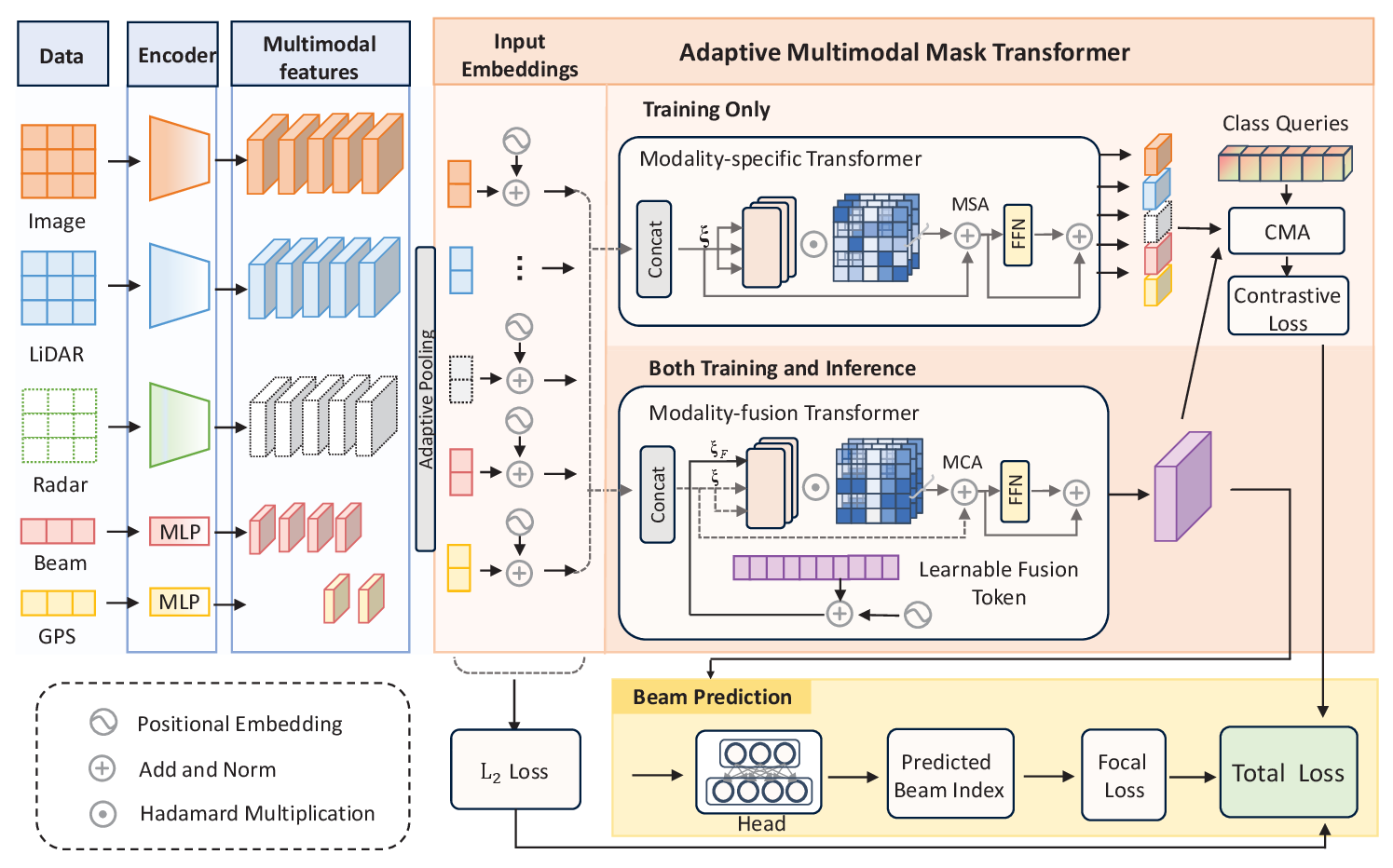}
\caption{Illustration of the proposed AMBER: The BS collects multimodal data, including images, LiDAR, radar, and GPS, along with historical beam indices. Modality-specific encoders are first employed to extract initial feature representations. Time-aware positional embeddings are then incorporated to capture temporal dependencies. The resulting features are fed into an adaptive multimodal mask transformer, which accommodates arbitrary missing modalities and extracts a unified fusion representation. Finally, the resulting fusion token is used to predict the optimal beam index.
 }   \label{fig_network_framework}
\end{figure*}

In this section, we introduce AMBER,  a framework  that jointly learns modality-specific features and aligns them within  a shared latent space to enable robust beam prediction under arbitrary missing-modality conditions.
AMBER employs cross-modality attention with a missing-modality-aware mask, allowing a learnable fusion token to adaptively query  information across the available modalities.
As illustrated in Fig. \ref{fig_network_framework}, AMBER consists of  three key components: (\textit{i}) modality-specific encoders, (\textit{ii}) an adaptive multimodal mask transformer incorporating a CMA module, and (\textit{iii}) a beam prediction head.

\subsection{Modality-Specific Encoders }

\textbf{Radar Encoder:}
For radar data processing, we adopt the method  in \cite{xu2023pidnet} to extract the range, angle and velocity information of the UE from the raw radar tensor ${\bf{X}}_{\text{R}}[t]\in \mathbb{R}^{M_R \times S_R \times A_R}$. Specifically,  range–angle (RA) and range–velocity (RV) maps are generated using two-dimensional fast Fourier transform (2D-FFT).
To ensure consistent spatial dimensions, the angle and velocity axes are zero-padded to a fixed size $N_{\text{FFT}}$ prior to performing the FFT. The RA and RV maps are computed as
\begin{align} \nonumber
  {\bf{X}}_{\text{R}}^{RA}[t] &=\left( \sum_{a=1}^{A_R}\left|\text{FFT}_{\text{2D}} ({\bf{X}}_{\text{R}}[t][:,:,a])\right|\right)^T, \\
  {\bf{X}}_{\text{R}}^{RV}[t] &= \sum_{a=1}^{M_R}|\text{FFT}_{\text{2D}} ({\bf{X}}_{\text{R}}[t][a,:,:])| .
\end{align}
Both ${\bf{X}}_{\text{R}}^{RA}[t]$ and ${\bf{X}}_{\text{R}}^{RV}[t]$ share the same spatial dimensions $S_R \times N_{\text{FFT}}$. To jointly preserve  spatial and velocity features without information loss, these two 2D maps are treated as single-channel images by introducing an explicit channel dimension (i.e., reshaped to $1 \times S_R \times N_{\text{FFT}}$). They are then concatenated along the channel dimension to form
\begin{align}
  \bar{\bf{X}}_{\text{R}}[t] &= \text{Concat}\left({\bf{X}}_{\text{R}}^{RA}[t], {\bf{X}}_{\text{R}}^{RV}[t]\right) \in \mathbb{R}^{2 \times S_R \times N_{\text{FFT}}}.
\end{align}
Next, min-max normalization is applied:
\begin{align}
  \tilde{\bf{X}}_{\text{R}}[t] = \mathcal{N}(\bar{\bf{X}}_{\text{R}}[t]),
\end{align}
where $\mathcal{N}(\cdot)$ denotes the  normalization operator defined as 
\begin{align}
  \mathcal{N}(\bf{X})= \frac{\bf{X}-\text{min}(\bf{X})}{\text{max}(\bf{X})-\text{min}(\bf{X})}.
\end{align}
The normalized tensor $\tilde{\bf{X}}_{\text{R}}[t]$ is then fed into a ResNet18-based encoder ${{\mathcal{P}}_{\text{R}}}$ \cite{he2016deep}, whose classification head
is removed and whose first convolutional layer is modified to accept the above two-channel radar input instead of the standard three-channel
RGB input.
Let ${{\mathcal{P}}_{\text{R}}}(\tilde{\bf{X}}_{\text{R}}[t])\in \mathbb{R}^{C \times H_{\mathrm{in}} \times W_{\mathrm{in}}}$ denote the output feature of the encoder, where
$C$, $H_{\mathrm{in}}$, and $W_{\mathrm{in}}$ represent the number of channels,
height, and width, respectively.
To obtain a fixed-size representation, an adaptive pooling operator
${\mathcal{P}}_{\mathrm{ada}}:
\mathbb{R}^{C \times H_{\mathrm{in}} \times W_{\mathrm{in}}}
\rightarrow
\mathbb{R}^{C \times V_A \times H_A}$
is then applied, where $V_A$ and $H_A$ represent the predefined output  dimensions.
The final radar encoder output is given by
\begin{align}
  {\bf{x}}_{\text{R}}[t] = {{\mathcal{P}}_{\text{ada}}}\left({{\mathcal{P}}_{\text{R}}}(\tilde{\bf{X}}_{\text{R}}[t])\right) .
\end{align}

\textbf{Image Encoder:} The image encoder extracts spatial and semantic information from the camera input. Each input image is first normalized using its mean ${\boldsymbol{\mu}}_{\text{I}}$ and standard deviation ${\boldsymbol{\sigma}}_{\text{I}}$, and then processed by a ResNet34 \cite{he2016deep} backbone followed by adaptive pooling. The image encoder input-output is defined as
\begin{align}
  {{\bf{x}}}_{\text{I}}[t] ={\mathcal{P}}_{\text{ada}}\left({\mathcal{P}}_{\text{I}}
  \left( \frac{{{\bf{X}}}_{\text{I}}[t] - {\boldsymbol{\mu}}_{\text{I}}}{{\boldsymbol{\sigma}}_{\text{I}}}\right)\right).
\end{align} 
Given the inherent visual complexity and rich semantic information present in camera images, a deeper neural network is essential for effective feature extraction. Consequently, we employ ResNet34 instead of ResNet18, as its deeper architecture provides the necessary representational capacity to adequately capture and discriminate these intricate features.

\textbf{LiDAR Encoder:} LiDAR data ${\bf{X}}_L[t] $ consists of raw 3D point clouds providing precise spatial geometry information. To obtain an efficient representation, the point cloud is projected onto a bird's-eye-view (BEV) grid of size $V_g \times H_g$ to form  a histogram ${\bf{X}}^{\text{H}}_L[t]$, where the maximum number of points per grid cell is capped at five to reduce outlier effects, and $V_g$ and $H_g$ are predefined constants to control grid size. The BEV representation preserves the spatial layout of the environment in a top-down perspective while compressing the 3D geometry into a computationally tractable 2D format.
The histogram is normalized and then fed into a modified ResNet18 backbone, where the first convolutional layer of ResNet18 is modified to accept input dimension, while the remaining residual architecture is kept unchanged and the classification head is removed. After adaptive pooling, the output is defined  as
\begin{align}
  {\bf{x}}_{\text{L}}[t] ={\mathcal{P}_{\text{L}}}\left(\mathcal{N}\left({\bf{X}}^{\text{H}}_L[t]\right)\right).
\end{align}

\textbf{GPS Encoder:} To better utilize GPS information, the GPS coordinates are  first transformed from longitude/latitude to  Cartesian coordinates relative to the BS, followed by min-max normalization. An MLP then encodes the positional  information as
\begin{align}
  {\bf{x}}_{\text{G}}[t] = {\mathcal{P}_{\text{G}}}\left(\mathcal{N}\left( {\bf{X}}_{\text{G}}[t]\right)\right).
\end{align}

\textbf{Beam index Encoder:} Historical beam indices are normalized and mapped via an MLP similar to the GPS encoder as
\begin{align}
  {\bf{x}}_{\text{B}}[t] = {\mathcal{P}_{\text{B}}}\left(\mathcal{N}\left({\bf{X}}_{\text{B}}[t]\right)\right).
\end{align}

\subsection{Adaptive Multimodal Mask Transformer }
Maintaining robustness to partial-modality inputs is crucial for real-world multimodal learning. To handle the incomplete modality availability, we propose an adaptive multimodal mask transformer, which is capable of learning multimodal representations directly from randomly incomplete inputs. Its core component  is a modality-aware mask that selectively blocks missing modality information during the attention process. This forces the model to focus exclusively  on the available data streams. For brevity, the time instance index $t$ is omitted in the following descriptions. 

The adaptive multimodal mask transformer consists of   input embeddings, the
modality-specific blocks,  the modality-fusion block, and a CMA module. The input embeddings transform modality specific features into token sequences. The modality-specific blocks independently process features for each modality,  and the modality-fusion block integrates  these single-modality representations into a a unified multimodal representation. Finally, the CMA module enables the fusion token to learn representations that are both discriminative and modality-cohesive.

\subsubsection{Input Embeddings}

After modality-sepcific encoding, the extracted  image/LiDAR/radar features  $\{{\bf{x}}_{\text{I}},{\bf{x}}_{\text{L}},{\bf{x}}_{\text{R}}\}$ are represented as nonoverlapping patches with dimensions  $C \times V_A \times H_A$. Together with GPS and beam index features, these patches are then transformed into a sequence of multimodal tokens. This transformation  employs spatial flattening followed by an MLP-based embedding layer, formulated as:
\begin{align}
  {\boldsymbol{\xi}} _i = \text{MLP}(\text{sf}({\bf{x}}_{i})), \;\;  i\in \mathcal I,
\end{align}
where
\begin{align}
  \{{\boldsymbol{\xi}} _{ \text{I}},{\boldsymbol{\xi}}_{\text{L}},{\boldsymbol{\xi}}_{\text{R}}\} \in \mathbb{R}^{ V_A H_A \times C},\;\; {\boldsymbol{\xi}}_{\text{B}} \in \mathbb{R}^{ (W-1)  \times C}, \;\; {\boldsymbol{\xi}}_{\text{G}} \in \mathbb{R}^{  2 \times C},
\end{align}
and $\text{sf}(\cdot)$ denotes spatial-flattening operator that collapses spatial dimensions $V_A$ and $H_A$ into a single dimension while preserving the channel dimension $C$. 
Due to the permutation-invariant nature of transformers, positional information is injected into the multimodal tokens using sinusoidal positional embeddings, defined as
\begin{align} \label{eq_pos_emb}
  {\bf{\text{PE}}}(k,2i) = \sin \left( \frac{k}{\Phi^{\frac{2i}{d}}} \right), \quad {\bf{\text{PE}}}(k,2i+1) = \cos \left(\frac{k}{\Phi^{\frac{2i}{d}}} \right),
\end{align}
where $k$ denotes  the position index, $i$ is the dimension index, and $\Phi$ is a scaling constant (typically 10000). 

For modalities with 2D spatial structure (image, LiDAR, radar), 2D positional embeddings are applied along the vertical ($k \in [0, V_A -1]$) and horizontal ($k \in [0, H_A-1$]) axes to preserve spatial geometry. 
To model temporal dependencies, a temporal-aware positional embedding is further added to each modality token, following the same sinusoidal form in \eqref{eq_pos_emb}. Specifically, $k$ ranges from 0 to $W-2$ for the beam index modality and from 0 to 1 for the GPS modality. By incorporating these spatial and temporal embeddings, we obtain the overall modality-specific token vector as 
\begin{align} \label{eq_modality_token_input}
  {\boldsymbol{\xi}}_ = [{\boldsymbol{\xi}}_{\text{I}}, {\boldsymbol{\xi}}_{\text{L}}, {\boldsymbol{\xi}}_{\text{R}}, {\boldsymbol{\xi}}_{\text{B}}, {\boldsymbol{\xi}}_{\text{G}}]. 
\end{align}

To capture the varying contribution of different modalities and handle missing-modality cases, we further introduce  a learnable 
modality-weight indicator ${\mathcal P}_{\text{ind}}$. Specifically,
${\mathcal P}_{\text{ind}}$ is parameterized by a trainable weight
vector $\mathbf{w}_{\text{mod}}
=[w_I,w_L,w_R,w_B,w_G]$, this vector is normalized
through a temperature-controlled softmax operation:
\begin{align}
  \boldsymbol{\alpha}
  =
  {\mathcal P}_{\text{ind}}(\mathbf{w}_{\text{mod}})
  =
  \text{Softmax}\!\left(\mathbf{w}_{\text{mod}}/\tau_{\text{mod}}\right).
  \end{align}
where $\tau_{\text{mod}}$ is the temperature hyperparameter. The
normalized weights satisfy $\alpha_i \ge 0$ and
$\sum_{i\in \mathcal{I}}\alpha_i = 1$.
Each modality token is then reweighted as
\begin{align}
{\boldsymbol{\xi}}_i =
\alpha_i \boldsymbol{\xi}_i,
\quad i\in\mathcal{I}.
\end{align} 
To avoid over-reliance on any single modality, particularly under random modality dropout, an $L_2$ regularization term is added to the loss function (formally introduced later in \eqref{eq_loss}) to promote balanced feature utilization. By applying the modality-availability vector $\bf{m}$ to filter out missing modalities, this regularization loss is formulated as 
\begin{align}
  L_2
  =
  \sum_{i\in\mathcal{I}} m_i\, w_i^2,
  \end{align}

\begin{figure}[t]
  \includegraphics[width=0.5\textwidth]{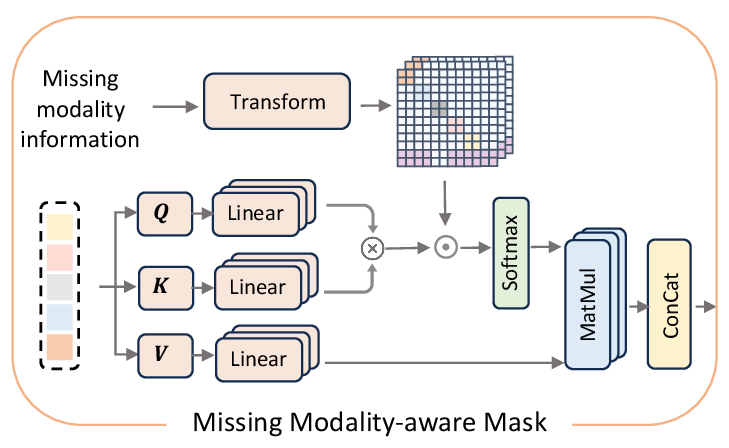}
  \caption{Illustration of the missing modality-aware mask within the multihead attention module. The mask is applied to the attention scores after the query-key multiplication, ensuring the fusion token only attends to available modalities. }\label{fig_mask_module}
\end{figure}

\subsubsection{Modality-specific Block}
The modality-specific block consists of a stack of transformer layers, each comprising a multi-head self-attention (MSA) module, a feed-forward network (FFN), and layer normalization (LN). For modality $j$, the MSA module projects the input token ${\boldsymbol{\xi}}_j$ into $H$ parallel heads. For the $h$-th head ($h=1,\dots,H$), the query, key, and value matrices are computed as
\begin{align} 
  {\bf{Q}}_{j,h} = {\boldsymbol{\xi}}_j {\bf{W}}_h^Q,\quad {\bf{K}}_{j,h} = {\boldsymbol{\xi}}_j {\bf{W}}_h^K, \quad {\bf{V}}_{j,h} = {\boldsymbol{\xi}}_j {\bf{W}}_h^V,
\end{align}
where ${\bf{W}}_h^Q, {\bf{W}}_h^K, {\bf{W}}_h^V \in \mathbb{R}^{C \times d_k}$ are learnable projection matrices for the $h$-th head, with $d_k = C/H$. The attention output for each head is calculated using scaled dot-product attention:
\begin{align} \label{eq_msa}
  \text{head}_{j,h} = \text{softmax}\left(\frac{{\bf{Q}}_{j,h} {\bf{K}}_{j,h}^T}{\sqrt{d_k}}\right) {\bf{V}}_{j,h}.
\end{align}
To facilitate efficient parallel computation across all modalities, we concatenate their input tokens into a unified sequence $\boldsymbol{\xi}$. Accordingly, a binary mask matrix ${\bf{M}}_{j,i}$ is designed based on $\bf{m}$ in \eqref{eq_modality_avail} to prevent noise propagation from missing modalities, which is illustrated in Fig. \ref{fig_mask_module}.
Specifically, ${\bf{M}}_{j,i}=1$ if modality $j$ can attend to  modality $i$, and ${\bf{M}}_{j,i}=0$ otherwise. Then, we have 
\begin{align}
  {\bf{M}}_{j,i} = \left\{ {\begin{array}{*{20}{c}}
    1,&{i = j}\\
    0,&{i \ne j}
    \end{array}} ,\right.
\end{align}
which indicates that the tokens from one modality can only attend to those from the same modality. 
Subsequently, each attention head's  output $\text{head}_{j,h}$ in \eqref{eq_msa} is  reformulated as
\begin{align} \label{eq_zj_cal}
  \text{head}_{j,h} =\sum_i^{} \left( \frac{ e^{\frac{{\bf{Q}}_{j,h} {\bf{K}}_{i,h}^T}{\sqrt{d_k}}}{ \odot {\bf{M}}_{j,i} }}{ \sum_{ i' } 
  e^{\frac{{\bf{Q}}_{j,h} {\bf{K}}^T_{i',h}}{\sqrt{d_k}}} \odot {\bf{M}}_{j,{i'}} } \right) {\bf{V}}_{i,h}.
\end{align}
Then, the outputs from all heads are concatenated and projected to form the final representation
\begin{align} 
  {\bf{Z}}_j = \text{MSA}({\boldsymbol{\xi}}_j) = \text{Concat}(\text{head}_{j,1}, \dots, \text{head}_{j,H}){\bf{W}}^O,
\end{align}
where ${\bf{W}}^O \in \mathbb{R}^{C \times C}$ denotes the output projection matrix, and ${\bf{Z}}_j$ represents the refined feature for the $j$-th modality.

This MSA module is followed by a residual connection and LN. The residual connection is employed to mitigate gradient vanishing and overfitting, while LN normalizes the feature across the feature dimension, reducing internal covariate shift and improving the training stability. The output of this first sub-layer is
\begin{align} \label{eq_ln_res}
  {\bf{Z}}'_j = \text{LN}( {\bf{Z}}_j + {\boldsymbol{\xi}}_j),
\end{align}
which is then processed by a FFN to further refine the contextual information and followed by a second residual connection and LN layer. The  output of the entire modality-specific block is
\begin{align} \label{eq_ffn}
  \bar{\bf{Z}}_j =\text{LN}( \text{FFN}({\bf{Z}}'_j)+{\bf{Z}}'_j). 
\end{align}

\begin{figure}[t]
  \includegraphics[width=0.45\textwidth]{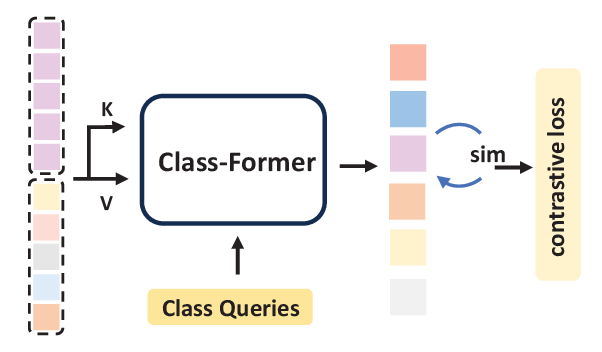}
  \caption{Illustration of the  structure of the CMA module.}\label{fig_class_former}
\end{figure}

\subsubsection{Modality-fusion Block}

As for modality-fusion block, it aggregates information from all modality-specific tokens using multi-head cross-attention (MCA). 
Specifically, a learnable fusion token ${\boldsymbol{\xi}}_{F}$, with  the same sequence length and feature dimension as the the summation of the modality-specific tokens, is further introduced and is randomly initialized.
It acts as a global aggregator, combining salient information from all available modalities, and enhances representation robustness under partial-modality conditions. 
Since it  lacks spatial and temporal embeddings, we add identical spatial and temporal embeddings to the fusion token as those added to their corresponding modality tokens.
The modality-fusion transformer's input is then represented  as 
\begin{align}
  \bar{\boldsymbol{\xi}} = [{\boldsymbol{\xi}}_{ \text{I}},{\boldsymbol{\xi}}_{\text{L}},{\boldsymbol{\xi}}_{\text{R}},{\boldsymbol{\xi}}_{B},{\boldsymbol{\xi}}_{\text{G}},{\boldsymbol{\xi}}_{\text{F}}] \in \mathbb{R}^{N \times C},
\end{align}
where $N= 2(3V_A V_H +W+1) $ is the  total number of tokens.

The fusion token ${\boldsymbol{\xi}}_F$ serves as the query, while the keys and values are derived from the modality-specific tokens ${\boldsymbol{\xi}}$. The cross attention computes the affinity between queries and keys, which is converted to the attention map through a spatial-wise softmax (operated along tokens), given as 
\begin{align}
  {\bf{Z}}'_F =\text{LN} ( \text{MCA} ({\boldsymbol{\xi}}_F,{\boldsymbol{\xi}} )+ {\boldsymbol{\xi}}_F) .
\end{align}
Within this block, we design a similar mask to ensure that the fusion modality $F$ can attend to itself and to any other available modalities, given as 
\begin{align}
  {\bf{M}}_{F,i'} = \left\{ {\begin{array}{*{20}{l}}
    1,&{\text{if modality}\; i' \;\; \text{is available} }\\
    0,&{ \text{otherwise }}
    \end{array}} ,\right.
\end{align}
where index $i'$ encompasses sensing modalities in \eqref{eq_modality_avail} and fusion modality $F$. 
The attention output follows an analogous form to \eqref{eq_zj_cal} and is omitted here for brevity.

The MCA also follows by FFN and LN, therefore, the output of the modality-fusion block is
\begin{align}
  \bar{\bf{Z}}_F = \text{LN}(\text{FFN} ({\bf{Z}}'_F)+{\bf{Z}}'_F),
\end{align}
where $\bar{\bf{Z}}_F$ is  a holistic summary of the entire multimodal input and is used for the downstream prediction task.

This design enables the model to dynamically adapt to varying modality availability while maintaining semantic integrity. Moreover, the temporal dependencies captured from past observations allow the model to leverage historical priors, improving robustness under sensor failure or data dropout.

\subsubsection{Class-former-aided Modality Alignment}

To enable the fusion token to learn both discriminative and modality-cohesive representations, we propose a CMA module, as shown in Fig. \ref{fig_class_former}. This module introduces a set of learnable class queries, ${\bf{c}}_j$ for each modality $j \in {\mathcal{I}}$ and a global query ${\bf{c}}_f$ for the fusion branch. These queries abstract  token-level features into higher-level class representations through cross-attention:
\begin{align}
  {\bf{c}}_j &= \mathrm{CA} ({\bf{c}}_j,{\bf{Z}}'_j), \;\;  j \in \mathcal{I}, \\
  {\bf{c}}_f &= \mathrm{CA} ({\bf{c}}_f,{\bf{Z}}'_F).
\end{align}
where $\mathrm{CA}(\cdot)$ denotes a standard cross-attention
operation, the first input is the query token and the second
input provides the key and value tokens.

Moreover, to align representations across modalities, we employ a contrastive learning loss, inspired by prior works \cite{chen2020simple,khosla2020supervised}. This loss maintains sample-level alignment while preserving modality-specific characteristics to avoid feature collapse. More specifically, it encourages similar (positive) pairs to be closer in the feature space,  while simultaneously pushing apart dissimilar (negative) pairs. We utilize cosine similarity as the distance metric, fostering a more nuanced alignment.
The contrastive loss is expressed as
\begin{align} \label{eq_conloss}
  L_c = \frac{1}{|\mathcal{A}|} \sum_{({\bf{c}}_j,{\bf{c}}_f,\{{\bf{c}}_i\}_{i=1}^{N} )\in \mathcal{A}  }^{} \left[ -\log \frac{e^{\text{sim}({\bf{c}}_j,{\bf{c}}_f)}/\tau}{ \sum_{i=1}^{N}e^{\text{sim}({\bf{c}}_j,{\bf{c}}_i)}/\tau } \right],
\end{align}
where we have  
\begin{align}
  \text{sim} ({\bf{c}}_i,{\bf{c}}_j) = \frac{ {\bf{c}}_i^T{\bf{c}}_j}{||{\bf{c}}_i||_2||{\bf{c}}_j||_2}, \nonumber
\end{align}
and $\tau$ is a temperature hyperparameter that scales the distribution of similarities. The set $\mathcal{A}$ represents the collection of all sample pairs, and $|\mathcal{A}|$ is its cardinality. For each token ${\bf{c}}_j$, the model wishes to correctly identify its positive sample  ${\bf{c}}_f$ among  one positive and remaining negative candidates  in the same mini-batch.
Minimizing this negative log-likelihood loss is equivalent to maximizing the mutual information between the representations of different modalities.

\textit{Remark}:  The modality-specific block and the CMA module are explicitly utilized during the training. Their primary role is to enforce cross-modality  alignment and serve as regularization constraints, thereby improving the quality and robustness of the learned representations and ultimately enhancing beam prediction performance. In contrast, the modality-fusion block, which yields the fusion features for beam prediction, is active during  both training and inference.

\subsection{Beam Prediction}
The fusion token output is processed by a beam prediction head, consisting of a dimensionality expansion layer followed by a fully connected network that outputs the predicted beam index $\hat{k}$. To further account for possible class imbalance in training dataset, we adopt  focal loss \cite{lin2017focal} to optimize the beam prediction performance, which is define as
\begin{align} \nonumber
  L_f = &- \hat{k}^* \beta_1 (1-\hat{k})^{\beta_2} \log(\hat{k})\\
  &- (1-\hat{k}^*)(  1-\beta_1) \hat{k}^{\beta_2} \log(1-\hat{k}),
\end{align}
where $\beta_1$ and $\beta_2$ are balance and scaling factors, respectively. The factor $\beta_1$ helps  mitigate the effect of class imbalance by assigning higher importance to the minority class. Furthermore, a larger 
$\beta_2$ increases the relative loss contribution of hard, misclassified examples while reducing the contribution from easy, well-classified ones. In addition, instead of using conventional one-hot encoding, we propose to utilize soft labels $\hat{k}^* $  to calculate the focal loss, where the values near the ground truth decay according to a Gaussian function across adjacent beam indices, enabling smoother supervision.

The overall training loss function is defined as
\begin{align} \label{eq_loss}
  L =  \lambda_f L_f + \lambda_c L_c + \lambda_r L_2,
\end{align}
where $\lambda_f$, $\lambda_c$  and $\lambda_r$ control the contribution of focal loss, contrastive loss, and L2 regularization loss, respectively.

\setlength{\tabcolsep}{10pt}
\begin{table}[t]
  \renewcommand{\arraystretch}{1.2}
	\begin{center}
		\caption{Scenarios on DeepSense 6G Dataset}
		\label{tab_Scenarios31_34}
		\begin{tabular}{c|ccc}
			\hline\noalign{\smallskip}
			\bf Scenarios & \bf Samples & \bf Street Scene &\bf Time Condition \\
			\noalign{\smallskip}
			\hline \hline
			\noalign{\smallskip}
      S31 & 7012 & Straight Road & Daytime \\
      S32 & 3235 & Turning Road & Daytime \\
      S33 & 3981 & Turning Road & Night \\
      S34 & 4439 & Crossroad & Night \\
			\hline
		\end{tabular}
	\end{center}
\end{table}

\section{Simulation Results} \label{sec_simulation}

\subsection{Simulation Settings}

\subsubsection{Dataset and training setup}

To evaluate the proposed AMBER framework under realistic mmWave communication conditions, we conduct experiments using the real-world DeepSense 6G dataset~\cite{alkhateeb2023deepsense}. To the best of our knowledge, this dataset is the first large-scale real-world sensing-communication dataset that provides synchronized multimodal sensing and communication measurements. Owing to its realism and comprehensive data modalities, it has been widely adopted as a representative benchmark for sensing-aided beam prediction tasks.

\setlength{\tabcolsep}{20pt}
\begin{table}[t]
  \renewcommand{\arraystretch}{1.2}
	\begin{center}
		\caption{System Parameter setting}
		\label{tab_system_parameter}
		\begin{tabular}{c|c}
			\hline\noalign{\smallskip}
			\bf Parameters & \bf Values\\
			\noalign{\smallskip}
			\hline \hline
			\noalign{\smallskip}
      mmWave Frequency & 60 GHz \\
      Number of BS Antennas & 16\\
      Codebook Size  & 64\\
      Sliding Window $W$ & 5 \\
			 Batch Size & 16 \\
       Learning Rate &  $1 \times 10^{-4}$ \\
       Traning Epochs & 20 \\
       Optimizer & AdamW \\
       Head Number $H$ & 8\\
       Dropout Rate & 0.1 \\
       Balance Factor $\beta_1$ & 0.25 \\
       Scaling Factor $\beta_2$ & 2 \\
       Weight of Beam Prediction Loss $\lambda_{f}$& 10 \\
      Weight of Contrastive Loss $\lambda_{c}$ & 0.2 \\
      Weight of Regularization Loss $\lambda_{r}$& 0.2 \\
			\hline
		\end{tabular}
	\end{center}
\end{table}
In this work, we use four urban street scenarios (Scenarios 31--34), as summarized in Table~\ref{tab_Scenarios31_34}, to comprehensively validate the robustness and effectiveness of AMBER under diverse real-world conditions. These scenarios cover different illumination and traffic environments, with Scenarios 31 and 32 collected during the daytime and Scenarios 33 and 34 collected at night. They also include heterogeneous road users, such as vehicles of different sizes and speeds as well as pedestrian activities, thereby introducing substantial environmental dynamics and sensing variability. These characteristics make DeepSense 6G a representative and challenging testbed for assessing the robustness and effectiveness of AMBER.

\setlength{\tabcolsep}{10pt}
\begin{table*}[t]
  \renewcommand{\arraystretch}{1.3}
	\begin{center}
		\caption{comparison of the proposed method with classical baselines based on average Top-3 Accuracy (\%) and DBA score}
		\label{tab_accuracy}
		\begin{tabular}{c|c|cccc|cc}
			\hline\noalign{\smallskip}
			\bf Modality & \bf Method & \bf S31 & \bf S32 & \bf S33 & \bf S34 & \bf Overall & \bf DBA \\
			\noalign{\smallskip}
			\hline \hline
			\noalign{\smallskip}
      GPS & Position-Aided-NN\cite{morais2023position} & 63.42 & 60.80 & 61.33 & 64.92 & 63.01 & 0.6667 \\
       & \bf AMBER(ours) & 97.57 & 83.20 & 80.31 & 73.74 & 82.08 & 0.8752 \\
       \noalign{\smallskip}
       \hline
      \noalign{\smallskip}
      Image+GPS & Vision-Position\cite{charan2022vision } & 90.38 & 78.23 & 83.32 & 80.14 & 80.64 & 0.8203    \\
      & TII-Transfuser  \cite{tian2023multimodal} & 88.32 & 79.92 & 76.63 & 73.56 & 75.82 & 0.7767 \\
      & BeMamba\cite{shi2025bemamba} & 99.57 & 83.86 & 83.89 & 82.73 & 85.96 & 0.8768 \\
      & \bf AMBER(ours) & 100.0 & 85.68 & 85.30 & 86.60 & 87.85 & 0.9203\\
      \noalign{\smallskip}
      \hline
      \noalign{\smallskip}
      Image+Radar & TII-Transfuser  \cite{tian2023multimodal} & 91.06 & 67.00 & 67.15 & 68.99 &  75.79 & 0.8904 \\
      & CMFE \cite{zhu2025advancing} & 100.0 & 72.75 & 71.44 & 74.73 & 76.64 & 0.9264\\
       & \bf AMBER(ours) & 100.0 &  87.61 & 86.16 & 87.04 & 88.25 & 0.9289\\
       \noalign{\smallskip}
      \hline
      \noalign{\smallskip}
      Image+Radar+GPS &  TII-Transfuser  \cite{tian2023multimodal} & 91.36 & 78.72 & 76.01 & 76.52 & 78.15 &0.8006 \\
      & BeMamba\cite{shi2025bemamba}  & 100.0 & 85.56 & 83.05 & 82.48 & 86.14 & 0.9008  \\
      & \bf AMBER(ours) & 100.0 & \bf 87.92 & 86.55 & 87.28 & 88.65 & 0.9234\\
       \noalign{\smallskip}
       \hline
      \noalign{\smallskip}
      Image+LiDAR+Radar+GPS & TII-Transfuser \cite{tian2023multimodal} & 98.29 & 82.37 & 83.47 & 79.56 & 83.12 &0.8573 \\
       & \bf AMBER(ours) & 100.0 & 86.81 & 86.07 & 87.42 & 88.63 & 0.9214 \\
       \noalign{\smallskip}
       \hline
      \noalign{\smallskip}
      Image+LiDAR+Radar+GPS+BeamIdx & Multimodal-LSTM \cite{shi2024multimodal} &  100.0 & 81.04 & 81.39 & 78.32 & 83.14 & 0.9040  \\
       & \bf AMBER(ours) & \bf 100.0 & 87.48 & \bf 86.26 & \bf 88.23 & \bf 89.07 & \bf 0.9294 \\
      \hline
      \noalign{\smallskip}
		\end{tabular}
	\end{center}
\end{table*}

\setlength{\tabcolsep}{10pt}
\begin{table}
  \renewcommand{\arraystretch}{1.2}
	\begin{center}
		\caption{Impact of Specific Missing Modality on Accuracy (\%)}
		\label{tab_Missing_Specific_Modality}
		\begin{tabular}{c|ccc}
			\hline\noalign{\smallskip}
			\bf Missing modality & \bf Top-1 &\bf Top-3 & \bf Top-5 \\
			\noalign{\smallskip}
			\hline \hline
      \noalign{\smallskip}
      Image & 61.06 & 86.25 &  93.08\\
      \noalign{\smallskip}
      LiDAR  & 63.88& 88.37 & 94.68 \\
      \noalign{\smallskip}
      Radar & 63.46 & 87.95 & 94.84 \\
      \noalign{\smallskip}
      Beam index  & 63.78 & 88.63 & 95.15\\
      \noalign{\smallskip}
      GPS & 63.81 & 88.81 &94.62\\
      \noalign{\smallskip}
      \hline
      \noalign{\smallskip}
      None &   \bf 64.15 & \bf 89.07 &\bf 95.33 \\
			\hline
		\end{tabular}
	\end{center}
\end{table}

For model training and evaluation, the dataset is randomly divided into 80\% training and 20\% testing subsets, each corresponding to an independent vehicle pass-by event with synchronized multimodal sensor data and beam index sequences. Ground-truth beam labels are obtained from the mmWave received power vectors at the BS. All experiments are conducted on a single NVIDIA RTX 4090 GPU using the PyTorch framework. The overall  simulation and training parameters are listed in Table \ref{tab_system_parameter}. AMBER is trained end-to-end using the loss function $L$ defined in \eqref{eq_loss}.
For radar and LiDAR feature extraction, we adopt a pretrained ResNet-18\cite{he2016deep}. For the image modality, we initialize the encoder with a pretrained ResNet-34\cite{he2016deep}, 
given that the image stream contains  richer semantic context and thus play a more critical role in accurate beam prediction (as analyzed in Section \ref{sec_performance}). Leveraging pretrained models can accelerate convergence and reduce overfitting by providing  well-regularized initial feature representations. 
The GPS encoder ${\mathcal P}_G$ is implemented as a three-layer MLP. Specifically, the normalized GPS input is first projected to 64 a 64-dimensional feature space via a linear layer, followed by LayerNorm and ReLU activation. The feature is then mapped from 64 to 256 dimensions, again followed by LayerNorm and ReLU. Finally, a linear layer projects the feature to the common embedding dimension of 256, followed by a final
LayerNorm. The beam encoder ${\mathcal P}_B$ adopts the same three-layer MLP architecture as ${\mathcal P}_G$, except that the input dimension of its first linear layer is adjusted to match the dimensionality of the beam features.

To emulate realistic sensor degradation, random modality masking is applied during training using predefined probabilities, thereby simulating partial-modality availability. The AdamW optimizer \cite{loshchilov2017fixing} with weight decay is used for all trainable parameters. A cosine learning-rate scheduler with five warm-up steps is adopted to avoid gradient explosion in the initial training phase. Specifically, the learning rate increases linearly during warm-up and subsequently decays according to a cosine schedule until reaching a near-zero minimum at the end of training. To further regularize the model, a dropout rate of 0.1 is applied within the attention and feed-forward layers of the transformer blocks.

\subsubsection{Evaluation Metrics}
Beam  prediction performance is evaluated using two complementary metrics:  Top-$K$ accuracy and distance-based accuracy (DBA) score. Specifically, the Top-$K$ accuracy measures whether the ground-truth beam index lies  within the $K$ predicted beam indices with the highest probabilities:
\begin{align}
\label{eq_topk}
\textbf{Top-K Accuracy} = \frac{1}{N}\sum_{i=1}^{N} {\bf{1}} \left\{y_i \in \mathcal{Q}_K \right\},
\end{align}
where $N$ is the total number of test samples, $y_i$ is
the ground-truth  label of the $i$-th sample, and $\mathcal{Q}_K$ is the set of top $K$ predicted beam indices.

The DBA score measures how close the Top-$K$ beam indices  are to the ground-truth beam index, which is defined as
\begin{align}
\label{eq_dba}
\textbf{DBA Score}= \frac{1}{K} \sum_{k=1}^{K} Y_k, 
\end{align}
where 
\begin{align}
  Y_k = 1 - \frac{1}{N} \sum_{i=1}^{N} \min_{ 1 \leq k' \leq k} \left[\min \left( \frac{1}{\Delta} | y_i - \hat{y}_{i,k'}  |  ,1 \right)\right],
\end{align}
and $\hat{y}_{i,k'}$ is the $k'$-th most-likely predicted beam  index for the $i$-th sample, while  $\Delta$ controls the normalization scale for beam proximity within the codebook.

\begin{figure}[t]
  \includegraphics[width=0.48\textwidth]{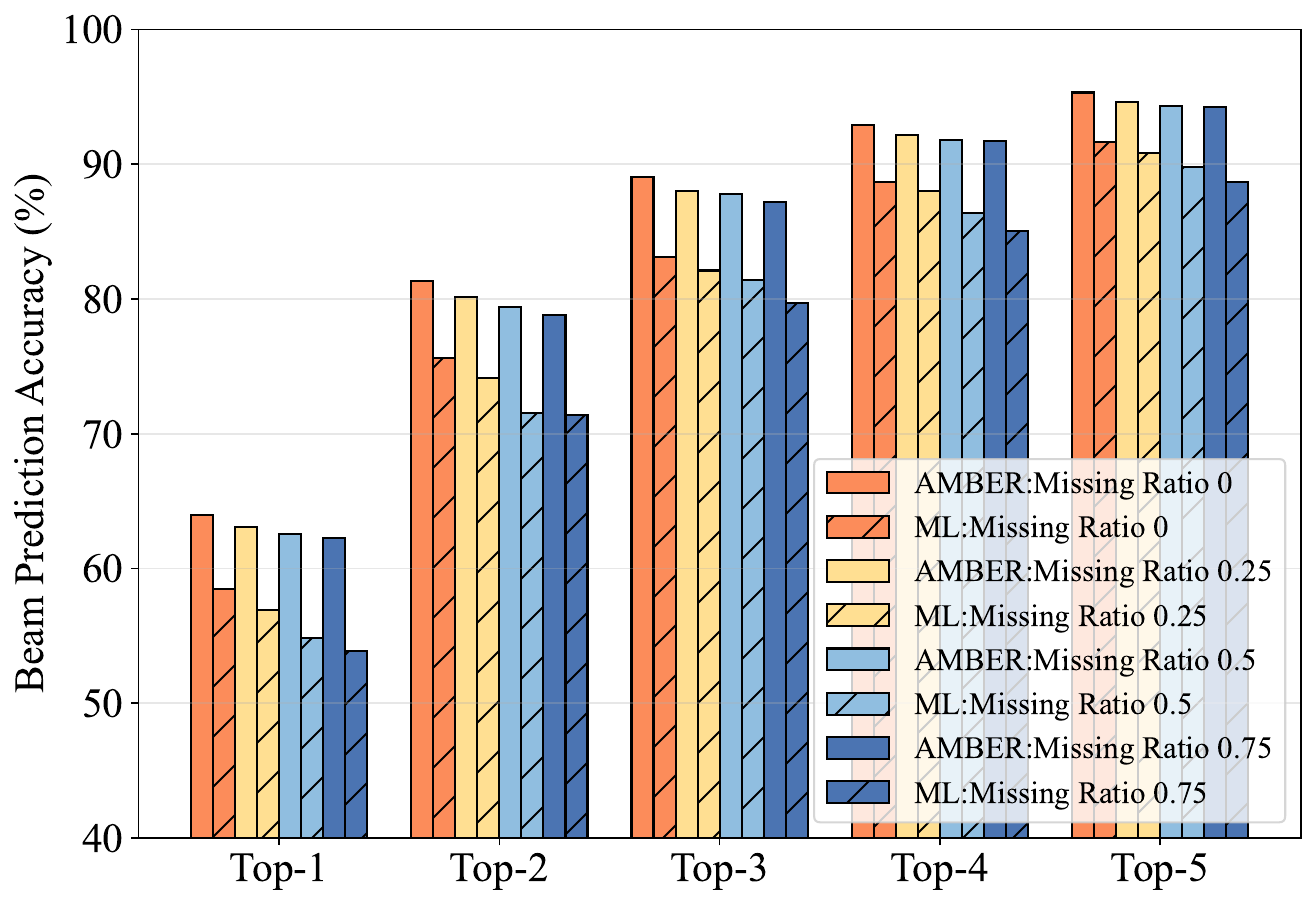}
\caption{ Beam prediction performance under different missing ratios when one modality is randomly missing, where the Multimodal-LSTM method is abbreviated as ML.}\label{fig_ChangeR1to025_Mis1}
\end{figure}

\begin{figure}[t]
  \includegraphics[width=0.48\textwidth]{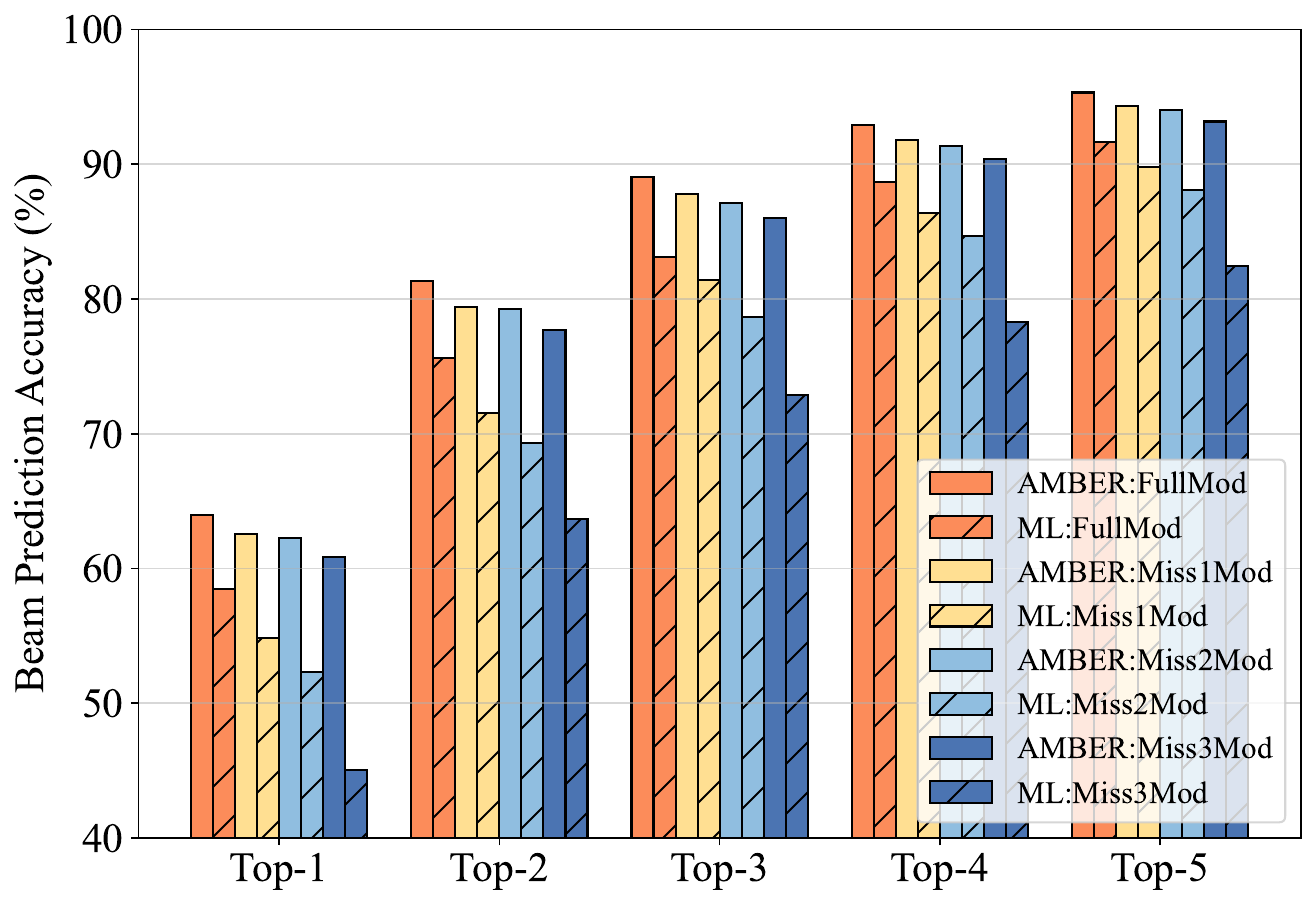}
\caption{Beam prediction performance under different missing modality number when randomly missing 50\% of the data, where the Multimodal-LSTM method is abbreviated as ML. And Miss$i$Mod indicates that $i$ modalities are missing.
}\label{fig_ChangeMod1to4_Mis05}
\end{figure}

\subsection{Performance Evaluation} \label{sec_performance}

Table \ref{tab_accuracy} presents  the performance comparison between the proposed AMBER method and several representative multimodal baselines in terms of Top-3 accuracy and DBA score. The baseline methods used in this study cover diverse modality fusion scenarios.
For AMBER, the same network architecture is used across all input settings, with only the modality mask adjusted to reflect the available sensing inputs.
In contrast, baseline models are individually designed and trained for each specific modality combination. 
The results demonstrate that AMBER consistently  achieves  state-of-the-art performance across  all metrics, verifying its effectiveness and its adaptability. Across all scenarios, AMBER achieves an average Top-3 accuracy of 89.07\% and a DBA score of 92.94\%. Specifically, when all modalities  are available, AMBER surpasses Multimodal-LSTM \cite{shi2024multimodal} by 5.93\% in Top-3 accuracy and 2.54\% in DBA score. Under more challenging nightime environments  (Scenarios 33 and 34), AMBER achieves 4.87\% and 9.91\% higher Top-3 accuracy than Multimodal-LSTM, highlighting  its strong robustness to degraded lighting and increased sensor noise.

To quantify the contribution of each sensing modality, we conduct a missing-modality analysis in which one modality is randomly removed during both training and inference.
The results, summarized in Table \ref{tab_Missing_Specific_Modality}, indicate that all sensing modalities contribute positively to beam prediction performance. Among them, the image modality proves most critical, contributing rich semantic cues and blockage information, while radar, LiDAR, and GPS inputs offer complementary  spatial and motion information that further enhance prediction accuracy.

The robustness of AMBER under partial-modality degradation is further assessed  by randomly masking one modality at varying ratios from 0\% to 75\%. As shown in Fig. \ref{fig_ChangeR1to025_Mis1}, Top-1 and Top-5 prediction accuracies gradually decrease as the missing ratio increases. Even under severe data loss, however, AMBER maintains over 62.3\% Top-1 and 94.3\% Top-5 accuracy, reflecting  strong resilience to missing sensory inputs.
Moreover, AMBER consistently outperforms the Multimodal-LSTM baseline across all masking ratios. Specifically, when the missing ratio increases from 25\% to 75\%, AMBER's Top-1 and Top-5 accuracies drop by only 1.67\% and 1.08\%, respectively, whereas Multimodal-LSTM suffers much larger drops of 4.64\% and 2.95\%. 
These results confirm AMBER's ability to effectively exploit the available data and maintain stable prediction performance under sensor failures or occlusions.

To further evaluate performance with multiple missing modalities, Fig. \ref{fig_ChangeMod1to4_Mis05} presents the results obtained by masking up to three modalities with 50\% masking ratio. While both methods achieve higher accuracy under a relaxed Top-$K$ criterion, AMBER consistently outperforms Multimodal-LSTM. As the number of missing modalities increases from 0 to 3, AMBER's Top-1 accuracy decreases moderately from 64.15\% to 60.86\%, whereas Multimodal-LSTM's exhibits a steep decline from 58.49\% to 45.03\%. The performance degradation rate of Multimodal-LSTM is approximately 4.2 times higher  than that of AMBER (13.46\% vs. 3.29\%), underscoring AMBER's reliability under incomplete modality conditions. This robustness is primarily attributed to the adaptive masking mechanism, which ensures that the transformer attends only to available modalities and avoids propagating noise from missing ones.

\setlength{\tabcolsep}{9pt}
\begin{table}[t]
  \renewcommand{\arraystretch}{1.2}
	\begin{center}
		\caption{Comparison of modality combination and computational complexity}
		\label{tab_complex}
		\begin{tabular}{c|ccccc}
			\hline\noalign{\smallskip}
			\bf Modality & \bf Top-1  &\bf Flops & \bf Latency \\
			\noalign{\smallskip}
			\hline \hline
      \noalign{\smallskip}
      BeamIdx+GPS  & 58.81\%  &  \bf 0.077G & \bf 7.98ms \\
      \noalign{\smallskip}
      BeamIdx+GPS+Radar & 59.85\% & 11.77G &  12.33ms \\
      \noalign{\smallskip}
      BeamIdx+GPS+LiDAR & 60.26\%  & 11.51G & 12.05ms  \\
      \noalign{\smallskip}
      BeamIdx+GPS+Image & 63.42\%  &24.14G & 15.58ms \\
      \noalign{\smallskip}
      BeamIdx+GPS+Image+Radar & 63.88\%  & 35.83G  & 19.03ms\\
      \noalign{\smallskip}
       Full Modality & \bf 64.15\%   &    47.27G &  22.64ms \\
			\hline
		\end{tabular}
	\end{center}
\end{table}

\begin{figure}[t]
  \centering
  \subfigure[Raw RGB image]{
      \includegraphics[scale=0.26]{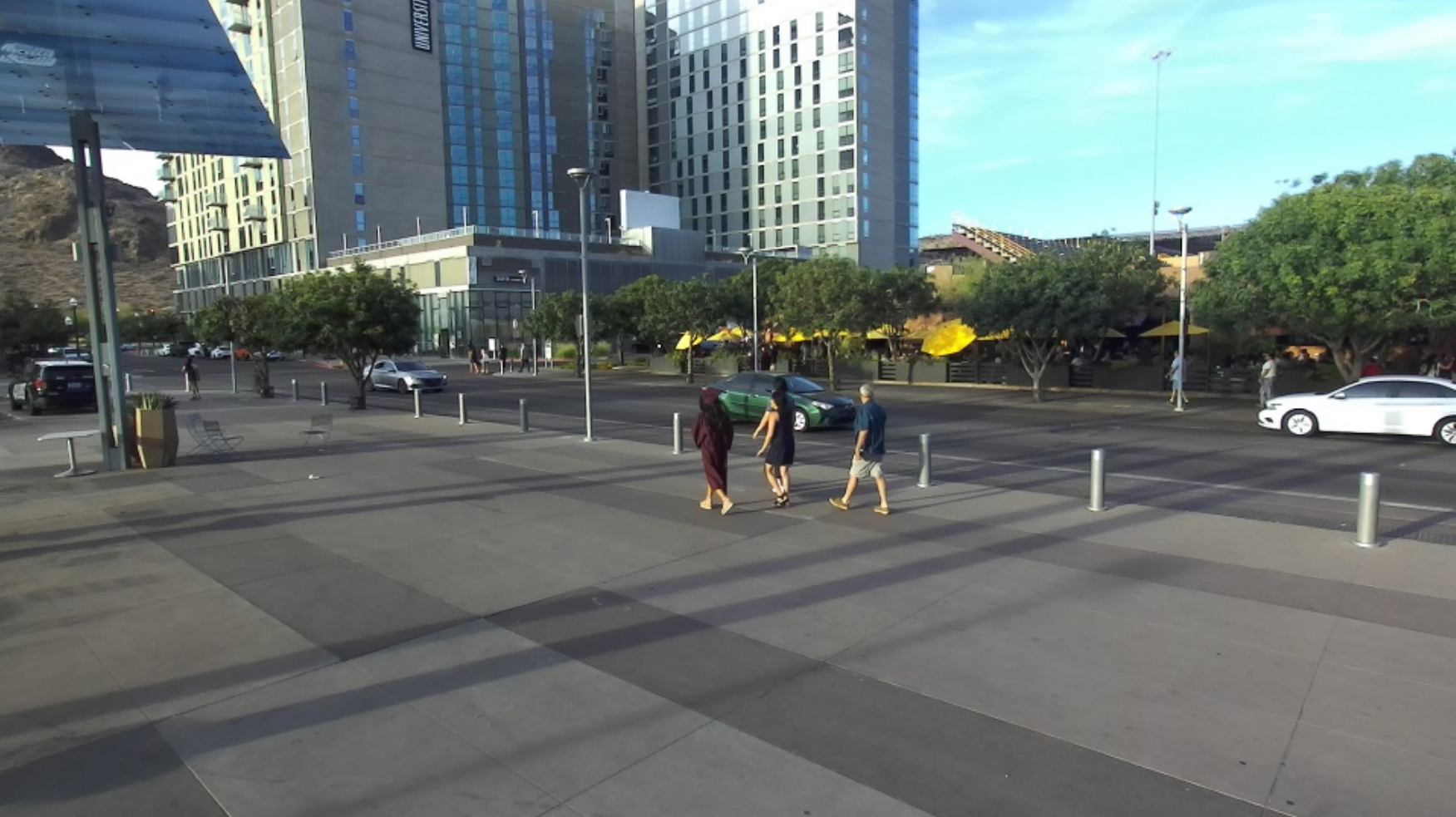}}

  \subfigure[Beam prediction result]{
      \includegraphics[scale=0.45]{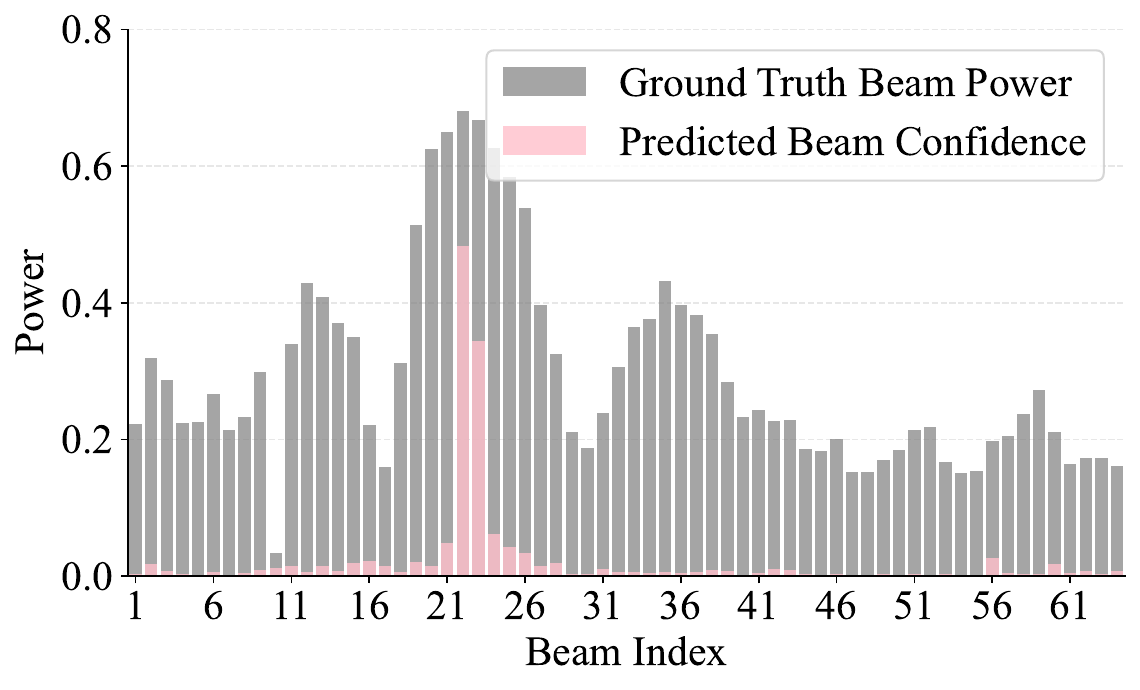}}     
  \caption{Visual example of beam prediction result. It describes the distributions of the ground truth beam power and the predicted beam confidence scores for 64 beam elements.}
  \label{fig_beam_distributions}
\end{figure}

We further report the inference FLOPs and inference latency under different modality combinations in Table \ref{tab_complex}. The full-modality configuration achieves the highest Top-1 accuracy of 64.15\%, but  also incurs the largest computational cost (47.27G FLOPs) and latency (22.64 ms). In contrast, lightweight configurations such as BeamIdx+GPS require negligible computation (0.077G) and achieve the lowest latency (7.98ms), at the cost of reduced  prediction accuracy. Among the evaluated settings, BeamIdx+GPS+Image provides a favorable accuracy-efficiency trade-off, achieving  63.42\% Top-1 accuracy while significantly reducing computational (24.14G FLOPs) and latency (15.58ms) compared to the full-modality setting. These results suggest that the proposed framework is not restricted to a fixed deployment mode, rather, it supports flexible modality combinations to accommodate diverse  vehicular hardware constraints and real-time requirements.
It is also observed that, although the theoretical FLOPs increase substantially, the measured inference latency grows more moderately. This discrepancy arises because practical GPU latency is influenced not only by arithmetic complexity, but also by factors such as memory access, kernel launch overhead, and parallel execution efficiency. Consequently, while FLOPs and latency exhibit consistent trends, they are not strictly proportional.

Fig. \ref{fig_beam_distributions} illustrates the beam prediction distribution across 64 beams in the codebook alongside  the corresponding RGB image. The pink bars (predicted confidence) align closely with the ground-truth beam (gray bars) and decay sharply across neighboring beams, demonstrating the effectiveness of the soft-label supervision strategy.

\begin{figure*}[t]
  \centering
  \subfigure[Overall training and validatoion loss]{
      \includegraphics[width=0.30\textwidth]{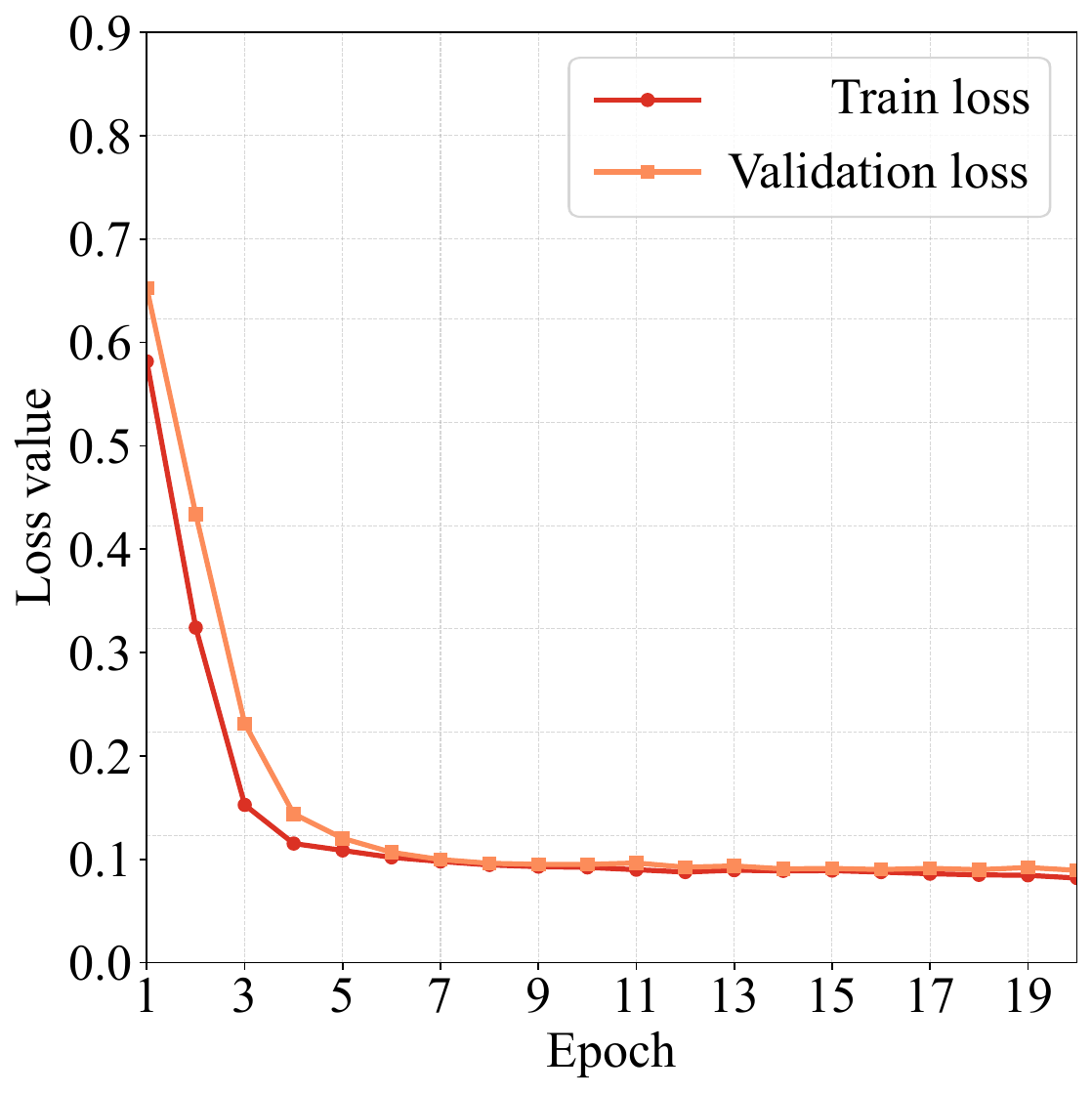}
      \label{fig_mic_train_valid_loss}
  } 
  \subfigure[Training accuracy of all scenarios in training set]{
      \includegraphics[width=0.30\textwidth]{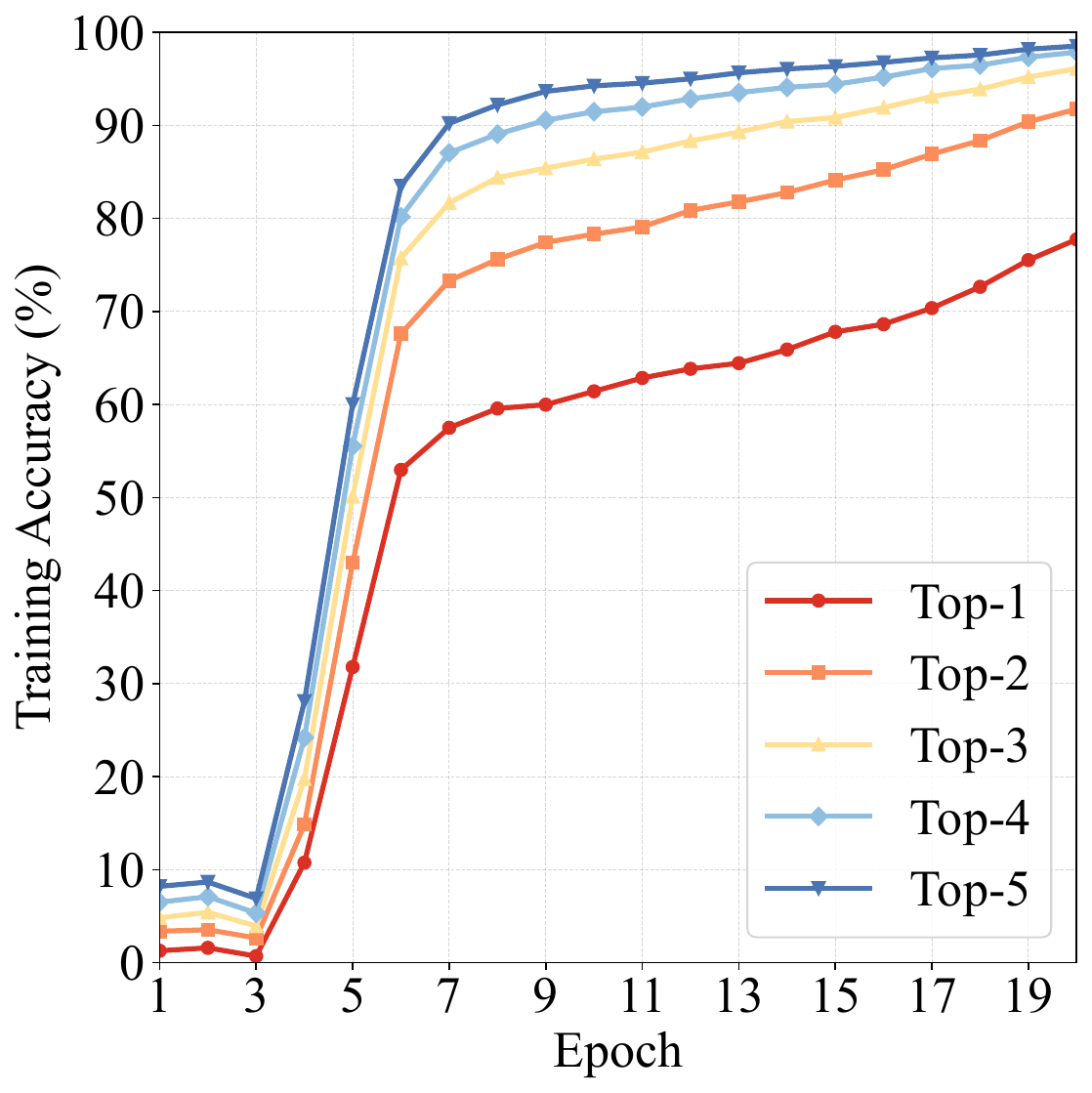}
      \label{fig_mic_train_accuracy}
  }
  \subfigure[Scenario 31: validation accuracy]{
      \includegraphics[width=0.30\textwidth]{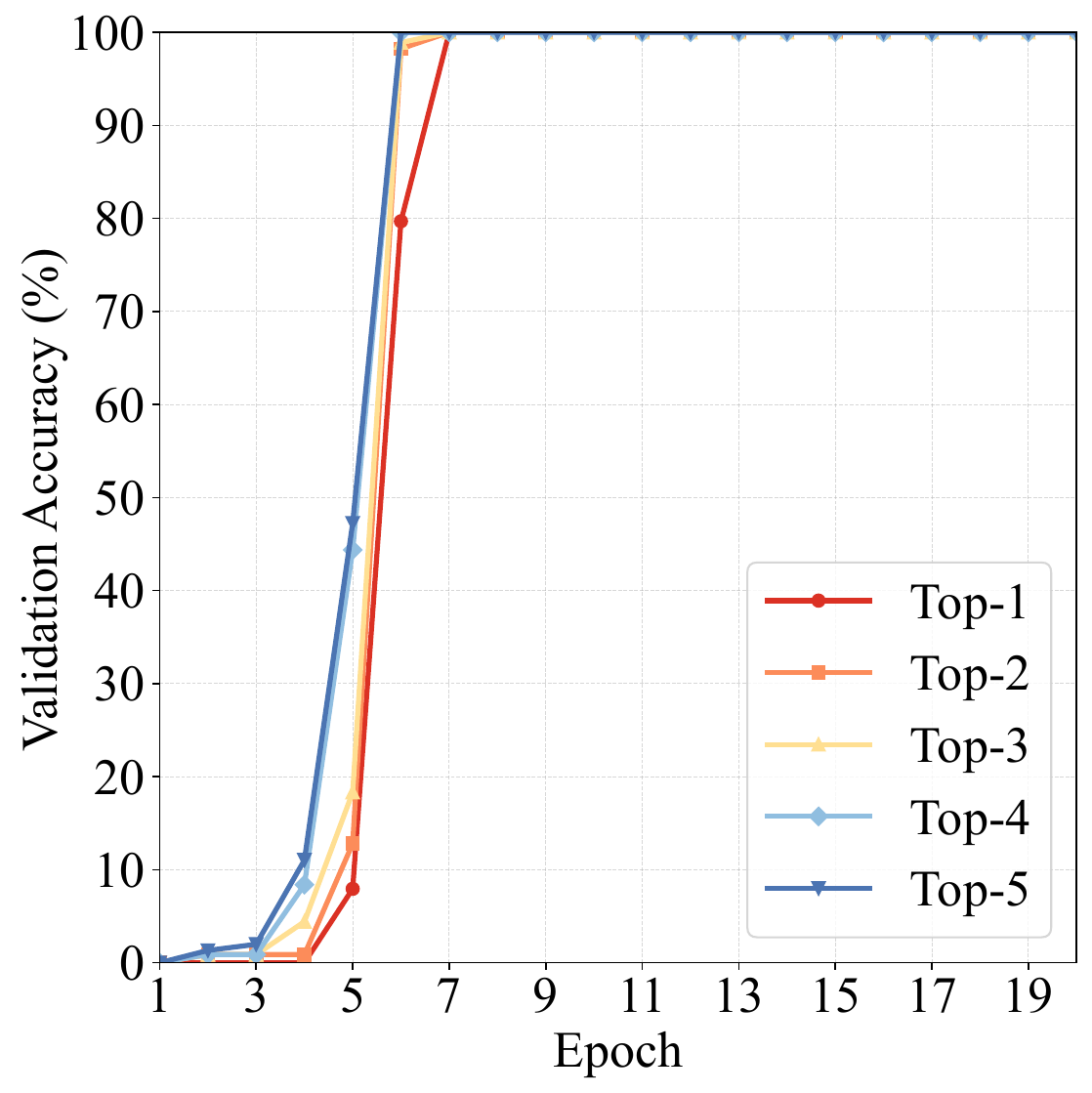}
      \label{fig_mic_valid_accuracy_31}
  }
  \subfigure[Scenario 32: validation accuracy]{
      \includegraphics[width=0.30\textwidth]{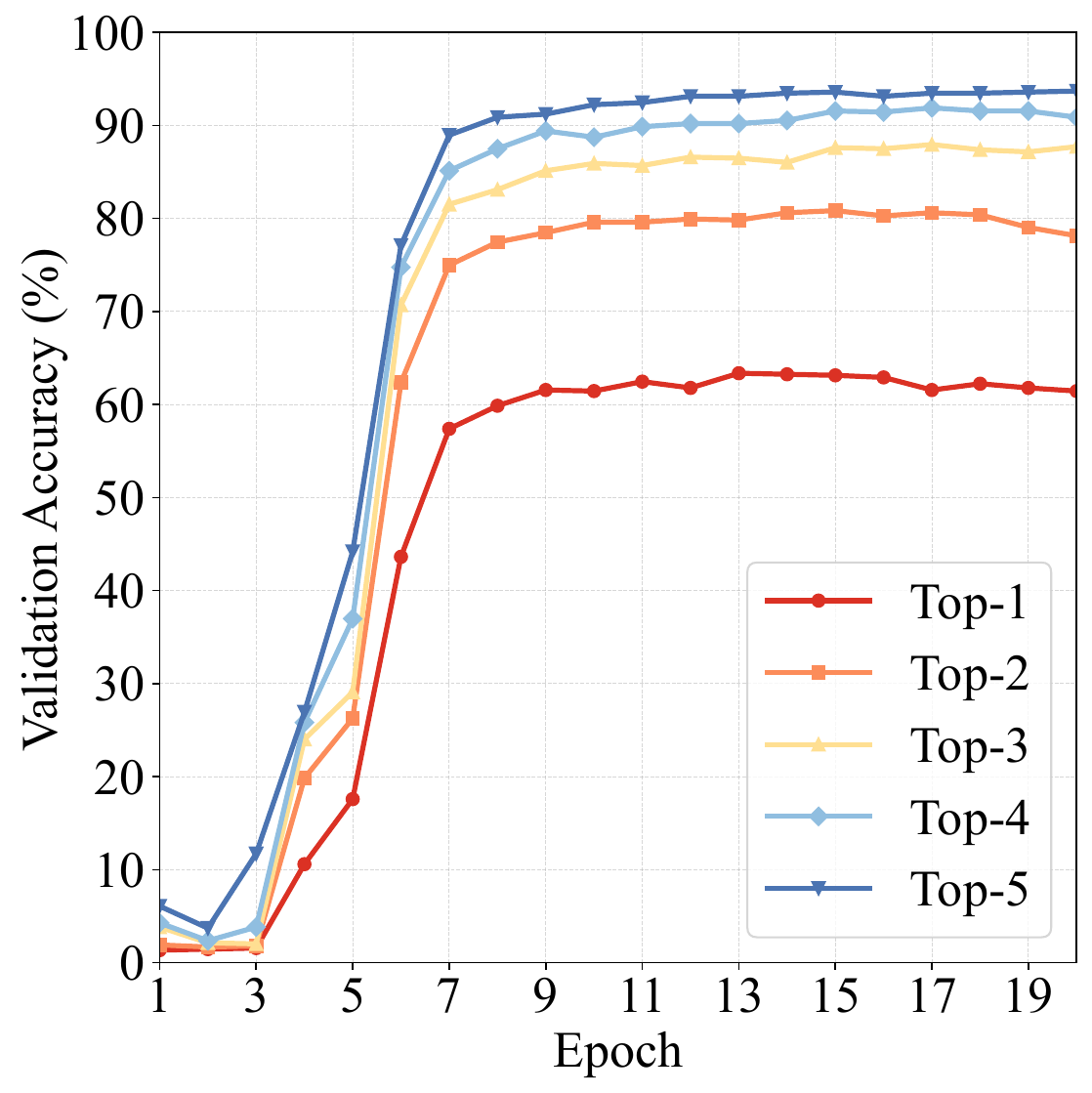}
      \label{fig_mic_valid_accuracy_32}
  }
  \subfigure[Scenario 33: validation accuracy]{
      \includegraphics[width=0.30\textwidth]{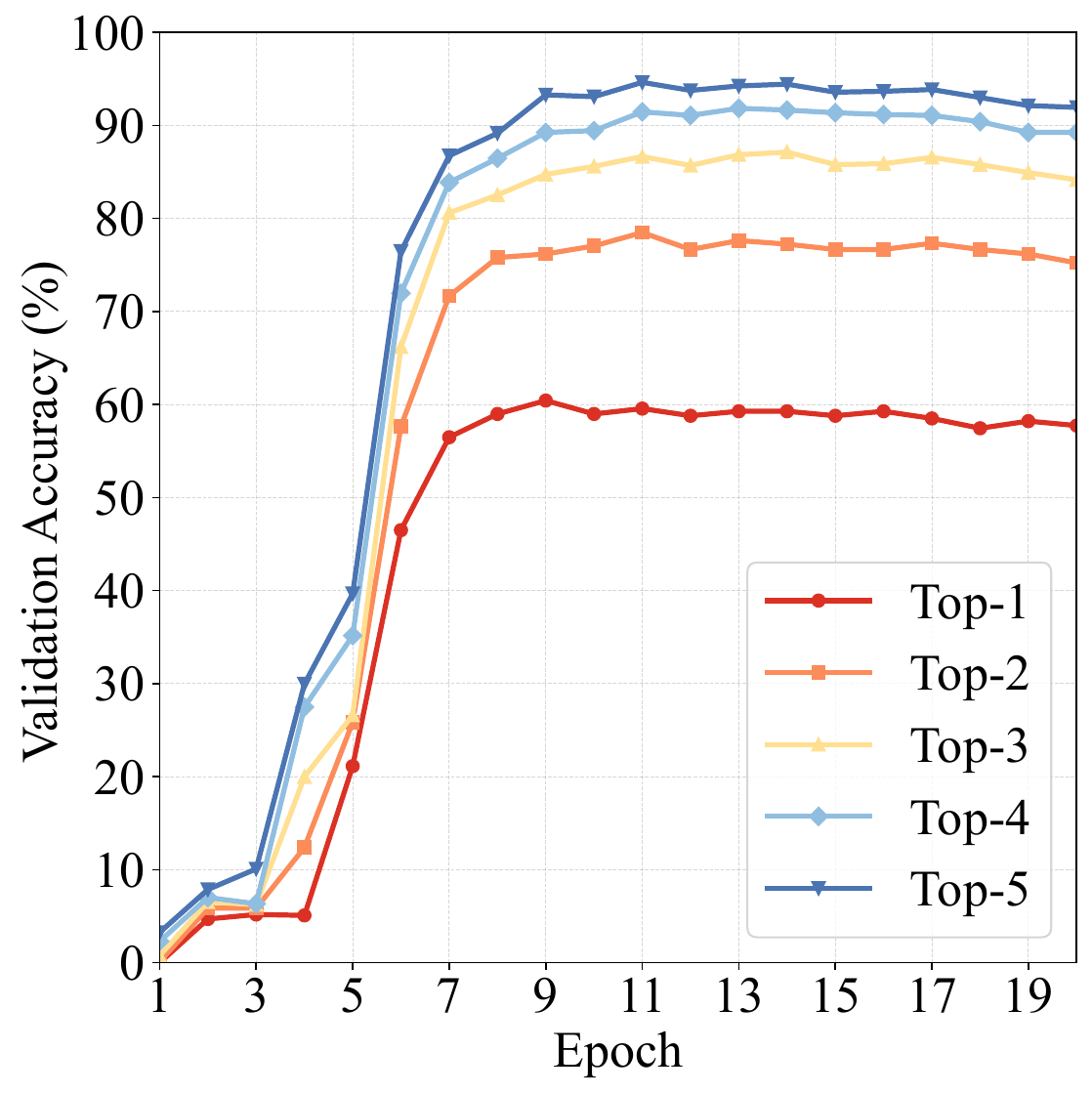}
      \label{fig_mic_valid_accuracy_33}
  }
  \subfigure[Scenario 34: validation accuracy]{
      \includegraphics[width=0.30\textwidth]{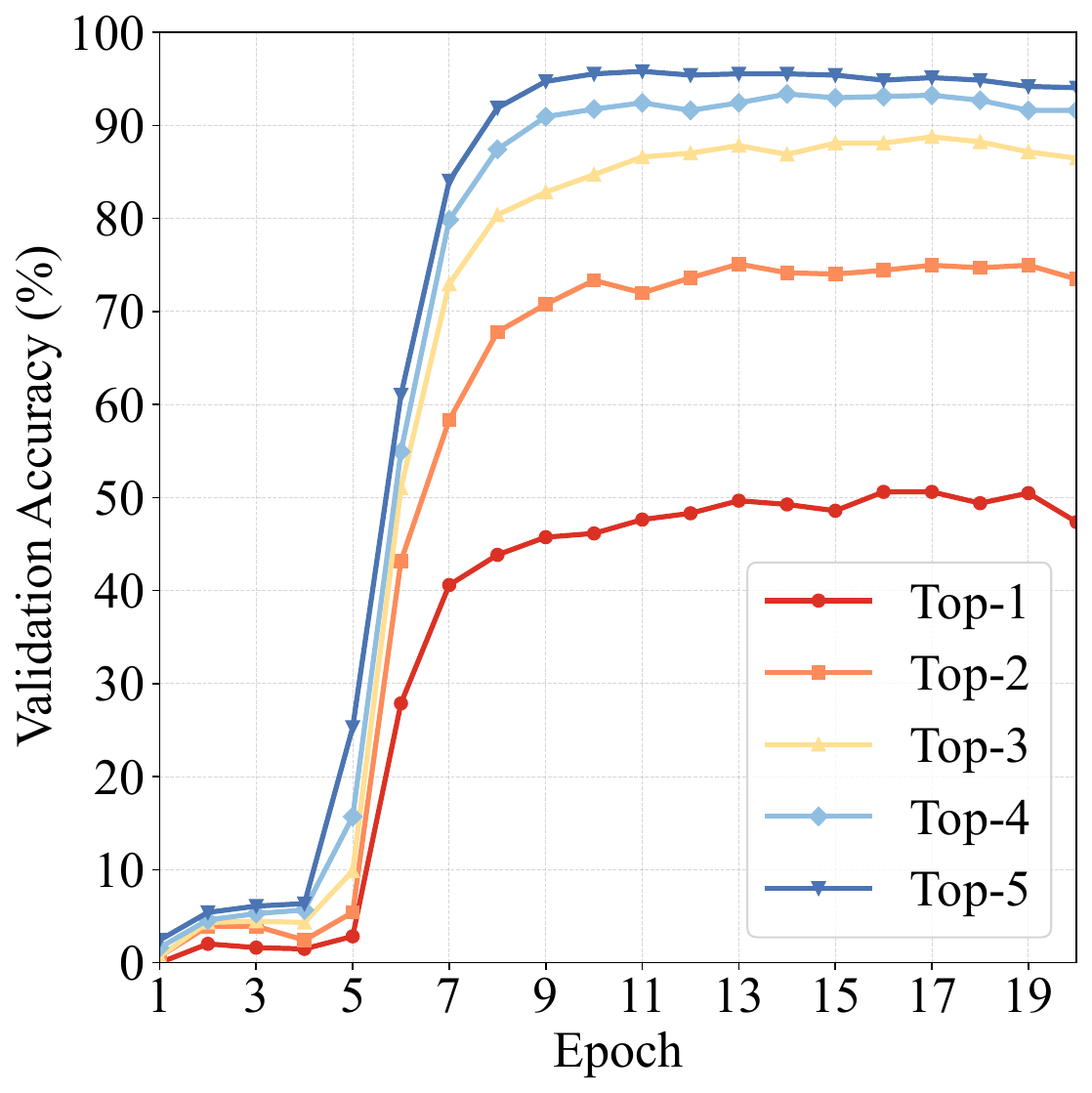}
      \label{fig_mic_valid_accuracy_34}
  }
  \caption{Convergence behavior of the training and validation loss, and the  training and validation accuracy for  31-34 scenarios of AMBER over 20 epochs.}
  \label{fig_fig_micPerMon}
\end{figure*}

Fig. \ref*{fig_fig_micPerMon} shows the training and validation behaviors  over 20 epochs across different scenarios. As shown in Fig. \ref{fig_mic_train_valid_loss}, both training and validation losses decrease rapidly and converge smoothly, indicating stable and well-conditioned optimization. Fig. \ref{fig_mic_train_accuracy} shows the evolution of training accuracy. During the initial 1-5 epochs, the learning-rate warmup phase introduces mild fluctuation, after which the accuracy improves steadily as the model learns discriminative representations.
The slower convergence of Top-1 accuracy relative to  Top-5 accuracy suggests that the model first captures coarse-grained beam structures before refining fine-grained distinctions. Figs. \ref{fig_mic_valid_accuracy_31}–\ref{fig_mic_valid_accuracy_34} further show that validation accuracies in all scenarios converge within nine epochs, demonstrating the efficiency and generalization capability of the proposed AMBER framework.

\subsection{Ablation Study}

In this subsection, we conduct an ablation study to quantify the contributions of several key components in AMBER, including the mask module, positional embedding, reweighting indicator, and cross-modal alignment module. Two incomplete-modality settings are considered: (i) {Miss1MR05}, where one randomly selected modality is missing with a masking ratio of 50\%; and (ii) {Miss2MR05}, where two randomly selected modalities are missing with the same masking ratio.
For clarity, we collectively refer to  the positional embedding, reweighting indicator, and CMA as robustness enhancement components (RCE). Although these components operate at different stages of the network, they are grouped together due to their shared objective of improving model robustness.

The ablation results are summarized in Table~\ref{tab_ablation}. Under the Miss1MR05 setting, introducing the mask module improves the Top-3 accuracy from 84.35\% to 87.15\%, corresponding to a gain of 2.80\%. 
Incorporating RCE further increases the Top-3 accuracy to 87.84\%, yielding an additional improvement of 0.69\%.  
Under the more challenging Miss2MR05 setting, the mask module provides an even larger Top-3 improvement of 3.83\%. These results indicate that the mask module contributes  the dominant performance gain and plays a key role in handling missing modalities, particularly as the degree of modality absence increases. In comparison, RCE serves as a complementary mechanism that further enhances feature representation and improves the robustness of multimodal fusion.

\setlength{\tabcolsep}{9pt}
\begin{table}[tbp]
  \renewcommand{\arraystretch}{1.2}
	\begin{center}
		\caption{Ablation study of beam prediction accuracy (\%).}
		\label{tab_ablation}
		\begin{tabular}{c|cc|ccc}
			\hline\noalign{\smallskip}
			\bf Modality & \bf Mask & \bf RCE & \bf Top-1 &\bf Top-3 & \bf Top-5  \\
			\noalign{\smallskip}
			\hline \hline
      \noalign{\smallskip}
       &   & & 60.38 & 84.35 & 93.20  \\
      \noalign{\smallskip}
      Miss1MR05 & \checkmark & & 62.44 & 87.15 & 93.97  \\
      \noalign{\smallskip}
       & \checkmark & \checkmark & \bf 63.13 &  \bf 87.84 & \bf 94.86 \\
      \hline 
      \noalign{\smallskip}
       &   & & 57.11 & 82.81 & 91.25  \\
      \noalign{\smallskip}
      Miss2MR05 & \checkmark & & 62.19 & 86.64 & 94.05\\
      \noalign{\smallskip}
       & \checkmark & \checkmark & \bf 62.89 & \bf 87.31 & \bf 94.62   \\
			\hline
		\end{tabular}
	\end{center}
\end{table}

\begin{figure}[t]
  \centering
  \subfigure[Model accuracy versus pooling size.]{
      \includegraphics[scale=0.4]{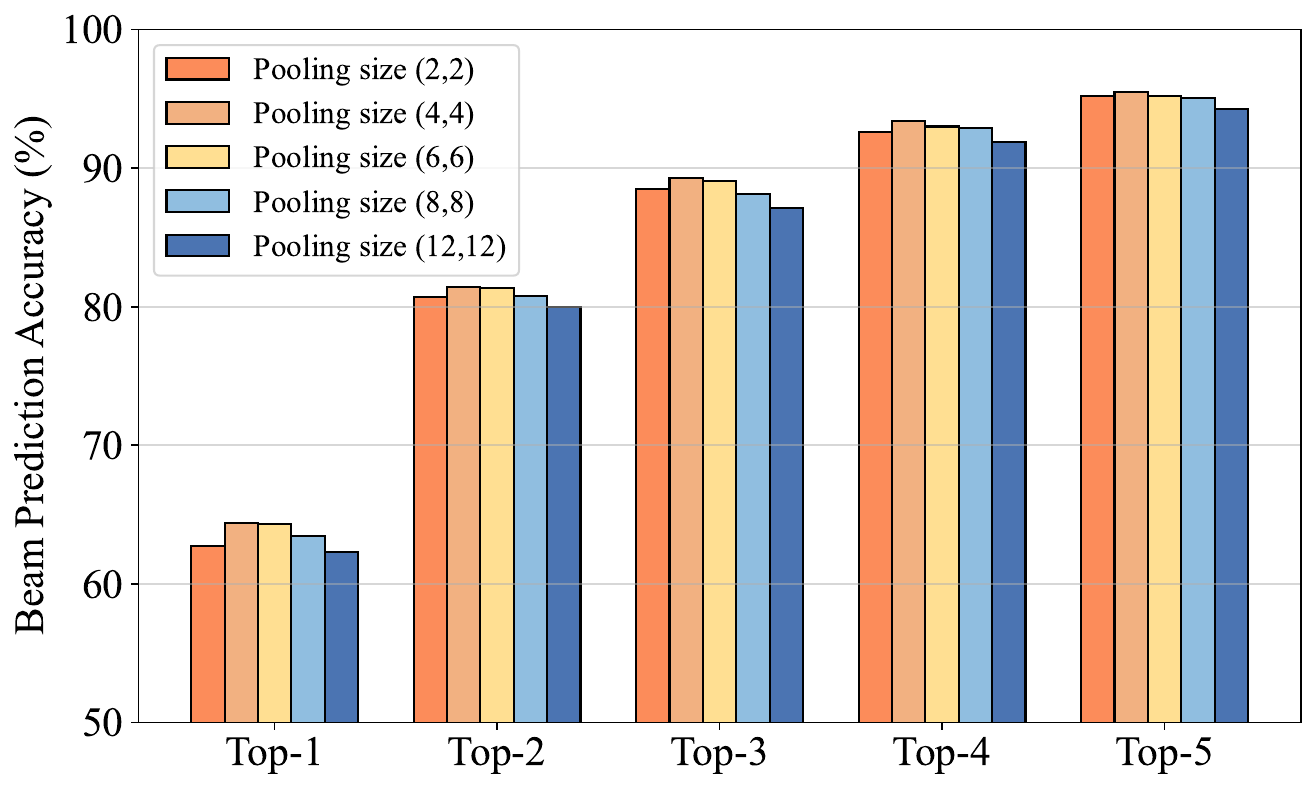}
      \label{fig_changePool}}
      
  \subfigure[Inference speed versus pooling size.]{
      \includegraphics[scale=0.41]{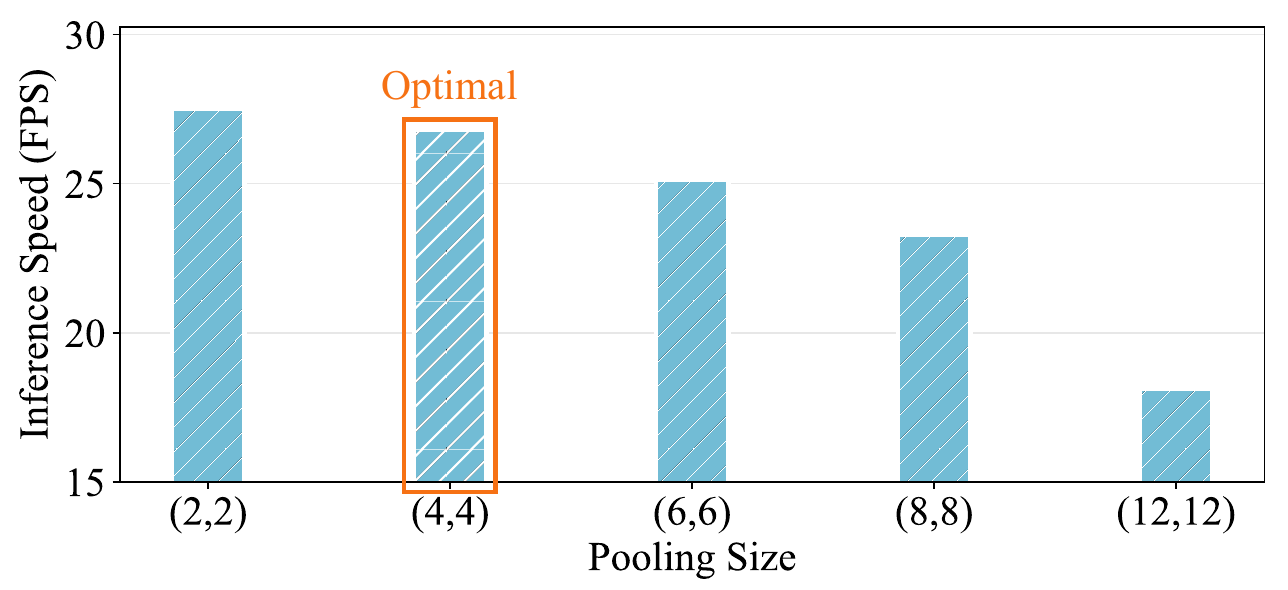}
      \label{fig_inference_speed}}

  \caption{The impact  of pooling size on model accuracy and inference speed: (a) shows the Top-$K$ accuracy achieved under  different pooling sizes; (b) illustrates the corresponding inference speeds.}
  \label{fig_pooling}
\end{figure}

The pooling size determines both the dimension of the modality-specific features (including LiDAR, radar, and camera features) and the length of the transformer input $\bar{\boldsymbol{\xi}}$. Its selection reflects a trade-off between model performance and inference efficiency.
The impact of the pooling size on accuracy and inference speed is analyzed in Fig. \ref{fig_pooling}. The 
accuracy initially improves as the pooling size increases but decreases beyond a certain point due to feature redundancy. Larger pooling sizes introduce excessive, repeated information, increasing the risk of overfitting and weakening generalization. 
Meanwhile, as shown in Fig.\ref{fig_inference_speed}, the overall runtime grows quadratically with approximately  the pooling size, since the number of input tokens scales accordingly. Among all evaluated configurations, a pooling size of (4, 4) achieves  the best trade-off, delivering  the highest Top-$K$ accuracy while maintaining an end-to-end validation throughput of 26.79 FPS. This throughput is measured over the complete validation pipeline, including data loading, CPU-to-GPU transfer, forward inference, and metric computation.

\section{Conclusion} \label{sec_conclusion}
In this work, we have proposed AMBER, a robust multimodal sensing-aided beam prediction framework designed to handle missing sensor inputs for mmWave  vehicular communication systems. AMBER integrates a missing-modality-aware mask Transformer and a learnable fusion token to intelligently accommodate incomplete sensory observations. In addition, a time-aware positional embedding and a CMA module were further introduced to enhance temporal coherence and semantic consistency across modalities. Extensive experiments verified that AMBER consistently outperforms existing approaches in Top-$K$ accuracy and DBA score, while maintaining strong robustness even when up to 75\% of modalities are missing. Moreover, AMBER effectively captures the temporal and semantic dependencies among multimodal features, enabling adaptive and reliable inference under degraded sensing conditions.

\ifCLASSOPTIONcaptionsoff
  \newpage
\fi

\bibliographystyle{IEEEtran}
\bibliography{IEEEabrv,refs.bib}

\end{document}